\documentclass[12pt,reqno]{amsart}

\usepackage{amsmath,amssymb}
\usepackage{graphicx}

\usepackage[english]{babel}
\newcommand{\be}{\begin{equation}}
\newcommand{\ee}{\end{equation}}

\begin{document}
\title{Moving bumps in theta neuron networks}% Force line breaks with \\

\author{Carlo R. Laing}
\email{c.r.laing@massey.ac.nz}
\address{School of Natural and Computational Sciences, 
Massey University, Private Bag 102-904 NSMC, Auckland, New Zealand. \\
phone: +64-9-414 0800 extn. 43512
fax: +64-9-4418136 }

\author{Oleh Omel'chenko}
\email{omelchenko@uni-potsdam.de}
\address{	University of Potsdam	,
Institute of Physics and Astronomy,
Karl-Liebknecht-Str. 24/25,
14476 Potsdam-Golm,
Germany}
\date{\today}
%\pacs{05.45.Xt, 05.45.Ac}

\begin{abstract}
%We consider large networks of theta neurons on a ring, synaptically coupled with an
%asymmetric kernel. Such networks support stable ``bumps'' of activity and we investigate
%the effects of the kernel asymmetry on the existence, stability and speed of these moving
%bumps using continuum equations formally describing infinite networks. 
We consider large networks of theta neurons on a ring, synaptically coupled with an asymmetric kernel.
Such networks support stable ``bumps'' of activity, which move along the ring if the coupling kernel is asymmetric.
We investigate the effects of the kernel asymmetry on the existence, stability and speed of these moving bumps using continuum equations formally describing infinite networks.
Depending on the
level of heterogeneity within the network we find complex sequences of bifurcations as the
amount of asymmetry is varied, in strong contrast to the behaviour of a classical neural
field model.

\end{abstract}

\maketitle

{\bf Classical 
neural field models of large scale activity in the cortex can sustain spatially-localised
``bumps'' of activity. These bumps move with a constant speed if the coupling kernel in the model
is asymmetric. We study such bumps in a ``next generation'' neural field model derived
from a network of theta neurons and find their behaviour to be very different from that in a classical model.
We find a complex sequence of saddle-node and Hopf bifurcations, resulting in multistability
and a nonmonotonic dependence of bump speed on the amount of asymmetry in the kernel.}

\section{Introduction}
Spatially-localised ``bumps'' of activity in neuronal networks have been studied
intensively over the past few 
decades~\cite{lai15,laitro02,coo05,brekil11,lai14a,pinerm01a}, as they are thought
to play a role in short term memory~\cite{bre12,Combru00} 
and the head direction system~\cite{zha96}, for example. In continuum neural field models
the most-studied systems are phenomenological~\cite{ama77} but recently it has been shown
how to derive a neural field model from an infinite network of theta 
neurons~\cite{lai14a,lai15,byravi19} (or equivalently, quadratic integrate-and-fire 
neurons~\cite{esnrox17}) by considering a spatially-extended version of the
networks studied by Luke at el.~\cite{lukbar13} and Montbri\'o et al.~\cite{monpaz15}. 
These derivations rely on the use of the Ott/Antonsen ansatz~\cite{ottant08,ottant09}
and assume that the neurons are heterogeneous, with the heterogeneous parameter being
distributed in a Lorentzian.

An obvious question is: do these ``next generation'' neural field models have any dynamics
which are not seen in classical neural field models? In terms of the
existence of stationary bumps, the answer seems to be ``no'' but next generation models
do support more interesting transient or time-dependent solutions than classical
ones. For example, Byrne et al.~\cite{byravi19} 
observed both oscillating bump solutions and travelling
waves, as well as more exotic solutions such as wandering bumps.
Esnaola-Acebes et al.~\cite{esnrox17}
found oscillatory transients which they referred to as ``synchrony-induced modes of oscillation''.
Schmidt and Avitabile~\cite{schavi19} studied ``oscillons'' --- 
spatially-localised time-periodic solutions caused
by either periodic temporal forcing or the interaction between an excitatory and an
inhibitory population. Lastly, Laing~\cite{lai17} showed that a transient {\em excitatory} stimulus
could switch a next generation model from a stable bump state to a stable quiescent state,
which it seems cannot occur in a classical model. 

Bump states are often studied on one-dimensional domains with periodic boundary conditions,
where the position of the bump encodes a periodic variable such as head direction~\cite{zha96}.
One is often interested in switching a network from a stable bump state to the stable 
``all-off'' state or vice versa using transient stimuli, but moving a bump along the domain
is also of interest. One way to do this is to break the symmetry of the coupling kernel, thereby
destabilising the stationary bump in favour of a bump which travels at constant 
speed~\cite{polngu16,zha96,xiehah02}. 

In this paper we consider breaking the symmetry of the coupling kernel in a neural field
model derived from a heterogeneous network of theta neurons, and investigate the effects
of this on the bumps' speed and structure. We find drastically different
behaviour from that found in a classical neural field model. Specifically,
we find (i) the speed of a bump 
is not a monotonic function of the amount of asymmetry, leading to possible bistability, (ii)
there may be Hopf bifurcations of travelling bumps, 
(iii) bumps may develop ``twists'' in their argument (when described using a complex
order parameter-like quantity)
corresponding to variations in instantaneous frequency as a bump passes a fixed position.

Our work is similar to that of~\cite{ome19,ome20}, although the model here has only one invariance ---
spatial translation --- while the model in~\cite{ome19,ome20} also has a global phase shift
invariance. Analysis of similar twisted states in heterogeneous networks of Kuramoto phase 
oscillators was presented in~\cite{omewol14}.
The outline of the paper is as follows. In Sec.~\ref{sec:model} we present the model
and briefly show some of the new types of solutions of interest. Sec.~\ref{sec:results}
contains the results of following bump solutions as  the asymmetry parameter is varied,
for various values of the neurons' heterogeneity. We conclude in Sec.~\ref{sec:disc}.

\section{Model}
\label{sec:model}
We consider the system of synaptically coupled theta neurons~\cite{lukbar13,lai14a,lai15,ermkop86}
\be
   \frac{d\theta_j}{dt}=1-\cos{\theta_j}+(1+\cos{\theta_j})(\eta_j+\kappa I_j); \qquad j=1,\dots N; \qquad \theta_j\in S^1,
   \label{Eq:Network}
\ee
where the input current to the $j$-th neuron includes a time-dependent term given by the sum
\be
   I_j(t) = \frac{2\pi}{N} \sum_{k=1}^N K_{jk} P_n(\theta_k(t))
   \label{Eq:InputSum}
\ee
and $P_n(\theta_k) = a_n ( 1 - \cos \theta_k )^n$ is the pulse-like signal
emitted by neuron $k$ when it fires ($\theta_k$ increases through $\pi$).
The positive integer parameter~$n$ controls the width of the pulse
and the constant~$a_n = 2^n (n!)^2 / (2n)!$ is determined
from the normalization condition $\int_0^{2\pi}P_n(\theta)d\theta=2\pi$.
The excitability parameters~$\eta_j$ are chosen
from a Lorentzian distribution with mean~$\eta_0$ and width $\gamma > 0$:
\be
   g(\eta)=\frac{\gamma/\pi}{(\eta-\eta_0)^2+\gamma^2},
\ee
and $\kappa$ is an overall coupling strength.
Moreover, we have $K_{jk}=K(2\pi|j-k|/N)$ where
\be
   K(x) = 0.1 + 0.3 \cos{x} + B \sin{x}
   \label{Eq:K}
\ee
is a $2\pi$-periodic coupling kernel.
For $B = 0$ the function~$K(x)$ is reflection symmetric
and has the Mexican-hat-like profile,
while for $B\ne 0$ it becomes asymmetric;
therefore we call~$B$ the {\it asymmetry parameter}.

\begin{figure}[t]
\includegraphics[scale=0.8]{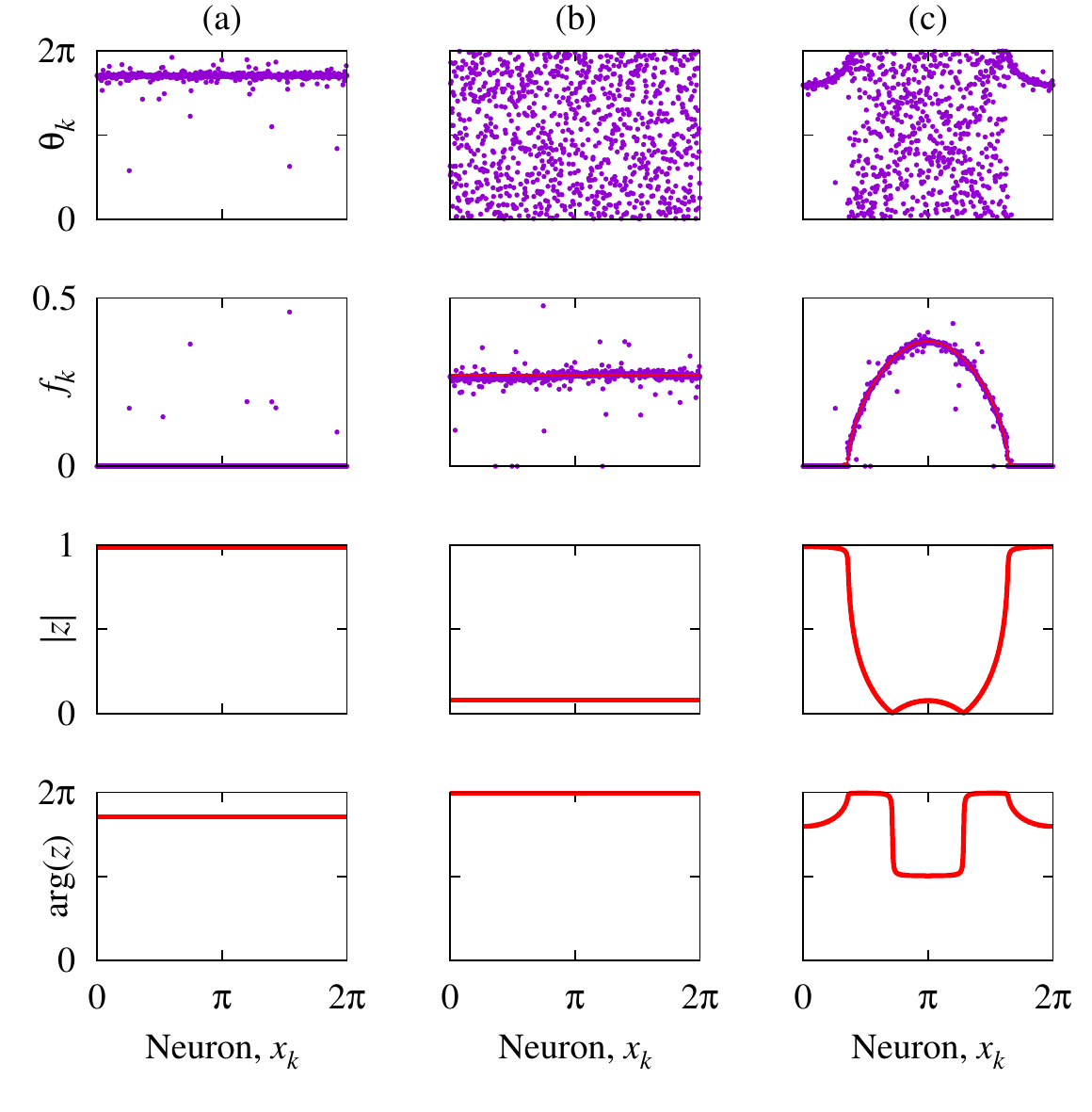}
\caption{(a), (b) Spatially uniform partially synchronized states
and (c) a bump state coexisting stably in system~(\ref{Eq:Network})--(\ref{Eq:InputSum})
for $\kappa = 2$, $\eta_0 = -0.4$, $n = 2$ and $\gamma = 0.01$.
The first and second rows show snapshots and mean frequencies (see (\ref{Def:f_k}))
of these states for theta neurons, while the third and fourth rows show
the moduli and the arguments of the corresponding stationary solutions~$z(x)$
to the neural field Eq.~(\ref{Eq:NeuralField}).
The thin lines in the second row show instantaneous frequencies computed by 
formula~(\ref{Formula:f}). Panel (a) shows a state in which most neurons are quiescent, with
mean frequency zero, whereas panel (b) shows a state in which most neurons are firing, with mean
frequency approximately $0.25$. In neither case does the mean frequency depend on spatial
position, unlike for the state shown in panel (c).}
\label{fig:introduction}
\end{figure}

A mean-field version of model~(\ref{Eq:Network})--(\ref{Eq:InputSum})
with constant kernel $K(x)$ was studied in~\cite{lukbar13}.
In~\cite{lai14a} one of the authors of this work derived
a neural field model describing the coarse grained
long-term dynamics of system~(\ref{Eq:Network})--(\ref{Eq:InputSum})
in the limit of infinitely many neurons, $N\to\infty$.
This dynamics is expressed in terms of the local mean field $z(x,t)$
representing the expected value of~$e^{i\theta}$ at~$x$ and~$t$.
The dynamics of local mean field~$z$ satisfies
\be
   \frac{\partial z}{\partial t} = \frac{(i\eta_0-\gamma)(1+z)^2 - i(1-z)^2}{2} + \kappa\frac{i(1+z)^2}{2} I
   \label{Eq:NeuralField}
\ee
where $i^2=-1$,
\be
   I(x,t) = \int_0^{2\pi} K(x-y) H(z(y,t);n) dy \label{eq:I}
\ee
is the input current at position~$x$ and
\be
   H(z;n)=a_n\left[C_0+\sum_{q=1}^n C_q(z^q+\bar{z}^q)\right]. \label{eq:H}
\ee
The coefficients~$C_q$ in~(\ref{eq:H}) are given explicitly by
\be
   C_q=\sum_{k=0}^n\sum_{m=0}^k\frac{\delta_{k-2m,q}(-1)^k n!}{2^k(n-k)!m!(k-m)!},
\ee
where $\delta_{i,j}=1$ if $i=j$ and $0$ otherwise.
From $z(x,t)$ one can calculate the instantaneous frequency of neurons
at position $x$ and time $t$ as the flux through $\theta=\pi$~\cite{lai17,monpaz15}:
\be
   f(x,t)=\frac{1}{\pi}\mbox{Re}\left(\frac{1-\bar{z}}{1+\bar{z}}\right).
\label{Formula:f}
\ee
In particular, for stationary solutions (when $z(x,t)$ and hence $f(x,t)$ do not depend on $t$)
formula (\ref{Formula:f}) yields also the central curve of the distribution of mean frequencies:
\be
f_k = \frac{1}{2\pi} \left\langle \frac{d\theta_k}{dt} \right\rangle
\label{Def:f_k}
\ee
where $\langle\cdot\rangle$ denotes time average.
In~\cite{lai14a} it was shown that for a certain range of widths~$\gamma$
the system~(\ref{Eq:Network})--(\ref{Eq:InputSum}) exhibits a bistable behaviour.
The ``all off'' state, where most of neurons are quiescent,
stably coexists with a so-called ``bump'' state,
where neurons in part of the domain are quiescent,
while neurons in the rest of the domain fire with mean frequencies which depend on their position.
For example, the ``all off'' state and the stationary bump state
for $\eta_0 = -0.4$, $n=2$, $\kappa=2$, $\gamma=0.01$
and $B = 0$ are shown in Figs.~\ref{fig:introduction}(a) and (c) respectively.
The corresponding mean fields~$z$ can be found as solutions to Eq.~(\ref{Eq:NeuralField})
with $a_n = 2/3$, $C_0 = 3/2$, $C_1 = -1$ and $C_2 = 1/4$.

By analogy with other phase models with broken reflection symmetry, we may expect
to find moving bumps in model~(\ref{Eq:Network})--(\ref{Eq:InputSum}) with $B > 0$.
\begin{figure}[t]
\includegraphics[scale=0.8]{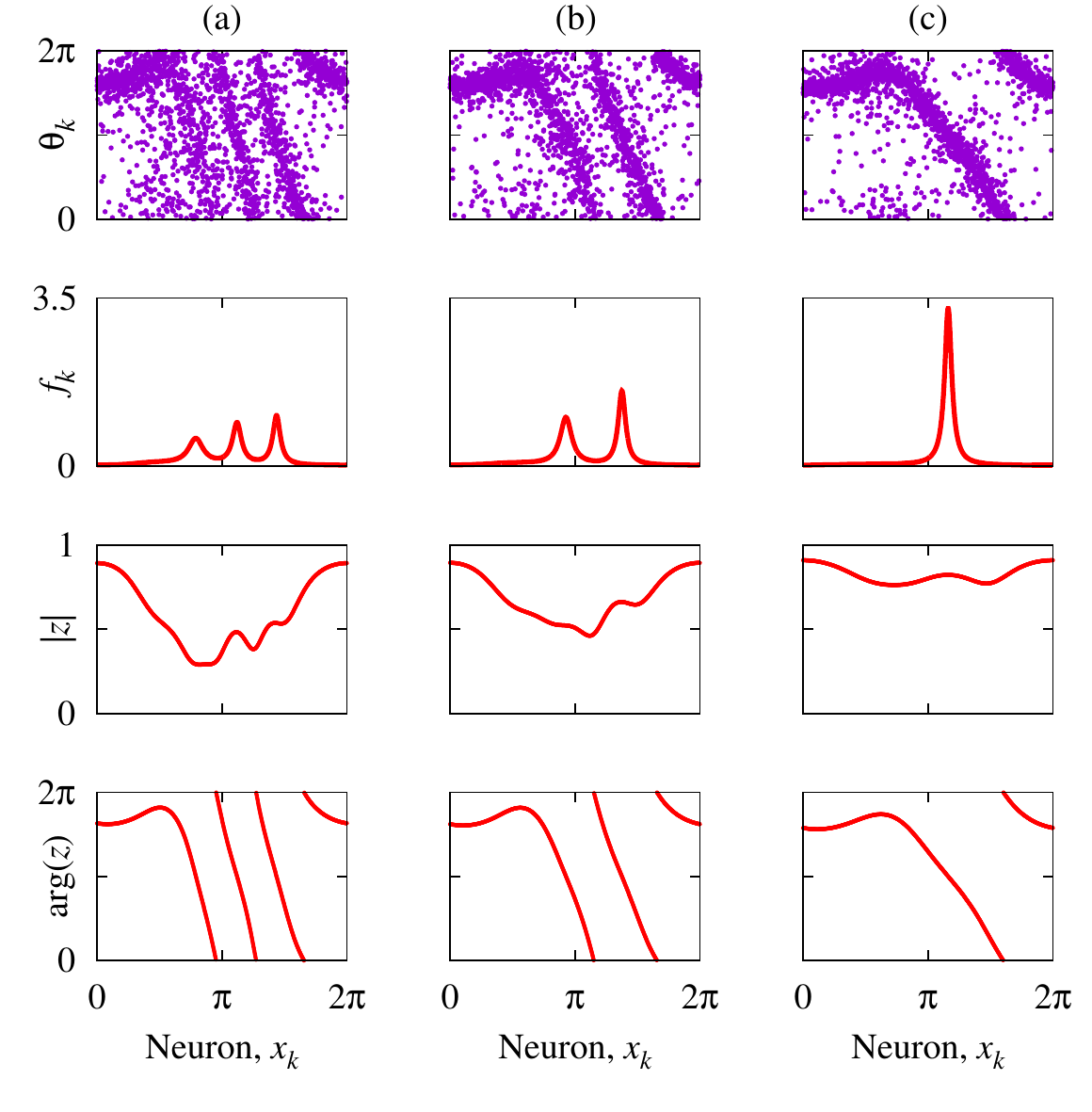}
\caption{Moving bumps in system~(\ref{Eq:Network})--(\ref{Eq:InputSum})
for (a) $B = 0.08$, (b) $B = 0.12$ and (c) $B = 0.16$.
All three are moving to the right.
Other parameters: $\kappa = 2$, $\eta_0 = -0.4$, $n = 2$ and $\gamma = 0.1$.
The first row shows snapshots of these states for theta neurons,
while the second, third and fourth rows show the instantaneous frequencies $f$ (see~(\ref{Formula:f})),
the moduli and the arguments of the corresponding travelling wave solutions~$z(x,t)$
to the neural field Eq.~(\ref{Eq:NeuralField}), where $s$ is the speed at which they travel.}
\label{fig:moving_bumps}
\end{figure}
It turns out that such bumps do exist, see Fig.~\ref{fig:moving_bumps},
but their behaviour is quite different from that in a classical neural field model
\be
   \frac{\partial u}{\partial t}=-u(x,t)+\int_0^{2\pi}K(x-y)F(u(y,t))dy \label{eq:NF}
\ee
where $F$ is some nondecreasing function (a sigmoid, or Heaviside) in at least three aspects:
(i) the speed is not a monotonic function of $B$, leading to possible 
bistability, (for~\eqref{eq:NF}, one can show that for $K(x)$ given by~\eqref{Eq:K},
the speed is $B/0.3$~\cite{polngu16}) (ii)
there may be Hopf bifurcations, and (iii) solutions may develop twists in their argument
corresponding to oscillations in average frequency as a bump passes a fixed position;
for example, moving bumps shown in Fig.~\ref{fig:moving_bumps} (a),(b) and (c)
are twist-$3$, twist-$2$ and twist-$1$, respectively. (The twist of a state is simply the net number
of multiples of $2\pi$ through which the argument of the complex quantity $z$ decreases
as the spatial domain is traversed once.)

\begin{figure}[t]
\includegraphics[scale=0.4,angle=270]{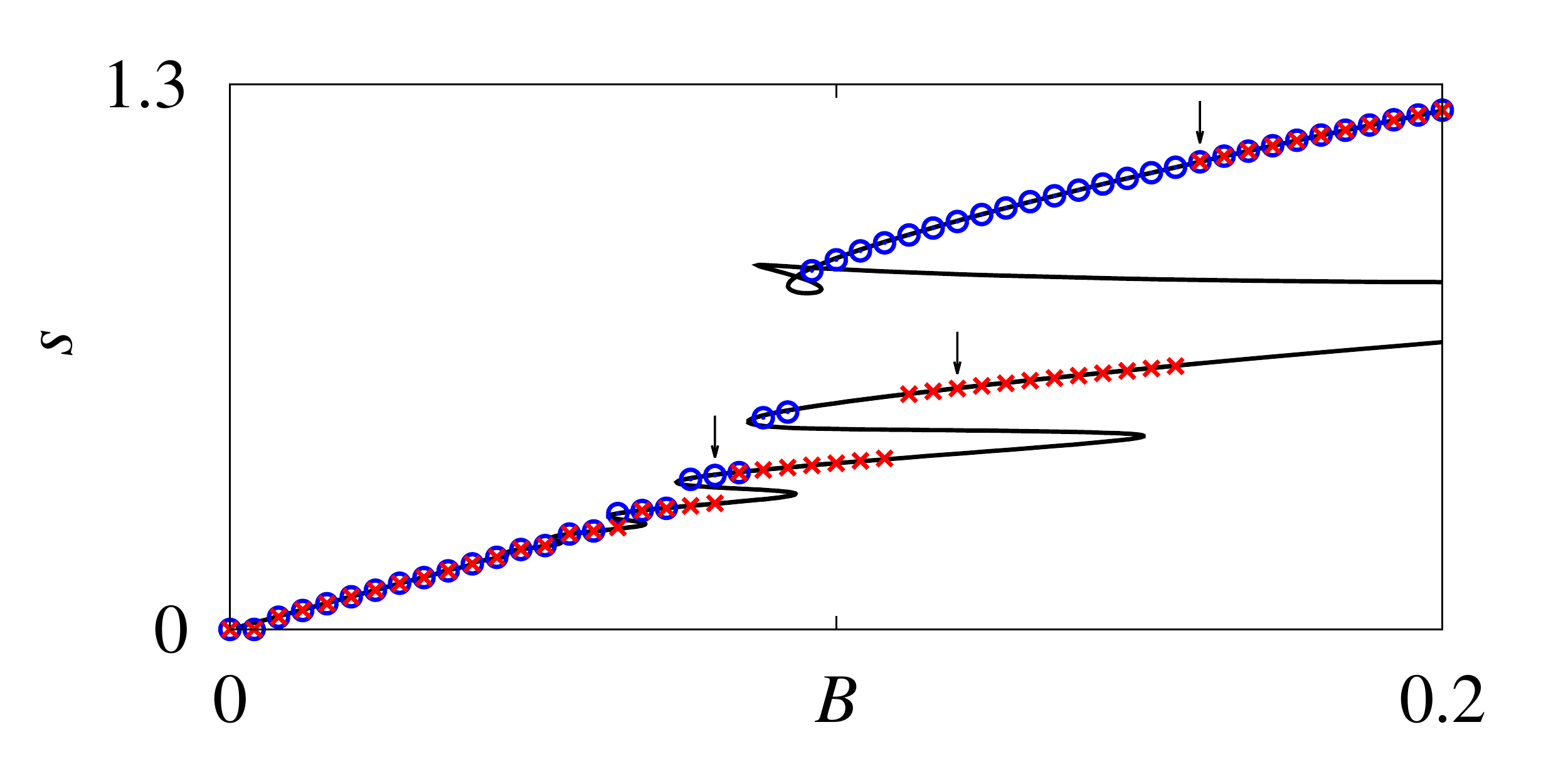}
\caption{Lateral speed~$s$ versus asymmetry parameter~$B$ for moving bumps
in system~(\ref{Eq:Network})--(\ref{Eq:InputSum}) with $N = 8192$ neurons.
Crosses and circles show lateral speeds obtained
in the forward and the backward $B$-sweeps, respectively (see details in the text).
Solid curve shows the lateral speed of travelling wave solutions
to the associated neural field Eq.~(\ref{Eq:NeuralField}) (see Fig.~\ref{fig:0_1a}, below).
Three vertical arrows indicate parameters of the travelling bumps
shown in Fig.~\ref{fig:moving_bumps} ($B=0.08,0.12$ and $0.16$).
Other parameters: $\gamma = 0.1$, $\kappa=2$, $\eta_0=-0.4$, $n=2$.}
\label{fig:scans}
\end{figure}

Stable moving bumps in system~(\ref{Eq:Network})--(\ref{Eq:InputSum})
can be studied by direct numerical simulations.
For this we used a Runge-Kutta integrator with the fixed time step $dt = 0.02$
and the data processing algorithm described in~\cite[Sec.~2.1]{ome20}
allowing us to extract the instantaneous position of a bump.
Taking a motionless bump as initial condition in system~(\ref{Eq:Network})--(\ref{Eq:InputSum})
with $N = 8192$ neurons we increased the asymmetry parameter
from $B = 0$ to $B = 0.2$ with the step $\Delta B = 0.001$.
For each value~$B$ we integrated system~(\ref{Eq:Network})--(\ref{Eq:InputSum})
over $2000$ time units and computed the mean lateral speed~$s$
from the last $1000$ time units of the trajectory.
The results of this forward $B$-sweep
are shown with crosses in Figure~\ref{fig:scans}.
Similarly, decreasing the asymmetry parameter from $B = 0.2$ to $B = 0$
we obtained a backward $B$-sweep shown with circles in Figure~\ref{fig:scans}.
Obviously, there are several hysteretic loops in the resulting diagram
indicating that the appearance of moving bumps
is mediated by a complicated bifurcation scenario.
It turns out that this scenario can be completely revealed
by path-following of spatially constant solutions
and travelling wave solutions to Eq.~(\ref{Eq:NeuralField}).

\section{Results}
\label{sec:results}
In this section we carry out a detailed analysis of the following solutions to Eq.~(\ref{Eq:NeuralField}):
\begin{itemize}
\item
spatially uniform states, i.e. fixed points $z(x,t) = a_0$ with constant profiles,

\item
stationary bumps, i.e. fixed points $z(x,t) = a(x)$ with $2\pi$-periodic non-constant profiles $a(x)$,

\item
moving bumps, i.e. travelling wave solutions of the form $z(x,t) = a(x - s t)$
where $a(x)$ is a non-constant bump profile and $s$ is a non-zero bump speed.
\end{itemize}
The numerical methods used are as follows. We discretised the domain $[0,2\pi]$ using
256 equally-spaced points and implemented the convolution in~\eqref{eq:I} using the fast
Fourier transform. Travelling wave solutions of~\eqref{Eq:NeuralField} with speed $s$
are steady states of
\be
   \frac{\partial a}{\partial t} = \frac{(i\eta_0-\gamma)(1+a)^2 - i(1-a)^2}{2} + \kappa\frac{i(1+a)^2}{2} I+s\frac{\partial a}{\partial x}.
   \label{Eq:NeuralFielda}
\ee
Having found a stable travelling wave through numerical integration of~\eqref{Eq:NeuralField}
and estimated its speed, we used this as an initial guess to find a steady state 
of~\eqref{Eq:NeuralFielda} using Newton's method. The spatial derivative was evaluated
spectrally. The stability of such a steady state is determined by the eigenvalues of the
linearisation of~\eqref{Eq:NeuralFielda} about the solution. A pinning condition
\be
   \int_0^{2\pi} \mbox{Re}[a(x)]\cos{(x)} \ dx=0
\ee
was appended to~\eqref{Eq:NeuralFielda} to remove the invariance under spatial shift of fixed
points of this equation.
Pseudo-arclength continuation was used to follow steady states of~\eqref{Eq:NeuralFielda}
as a parameter was varied~\cite{lai14b,doekel91,gov00} and the twist of a state was
determined by measuring the net variation in the phase of $z$ over the spatial domain using
the Matlab command {\tt unwrap}.
Several computations were repeated with 512 spatial points rather than 256 and no differences
in results were observed.

\subsection{Symmetric coupling kernel $B = 0$}

\begin{figure}[t]
\includegraphics[scale=0.8]{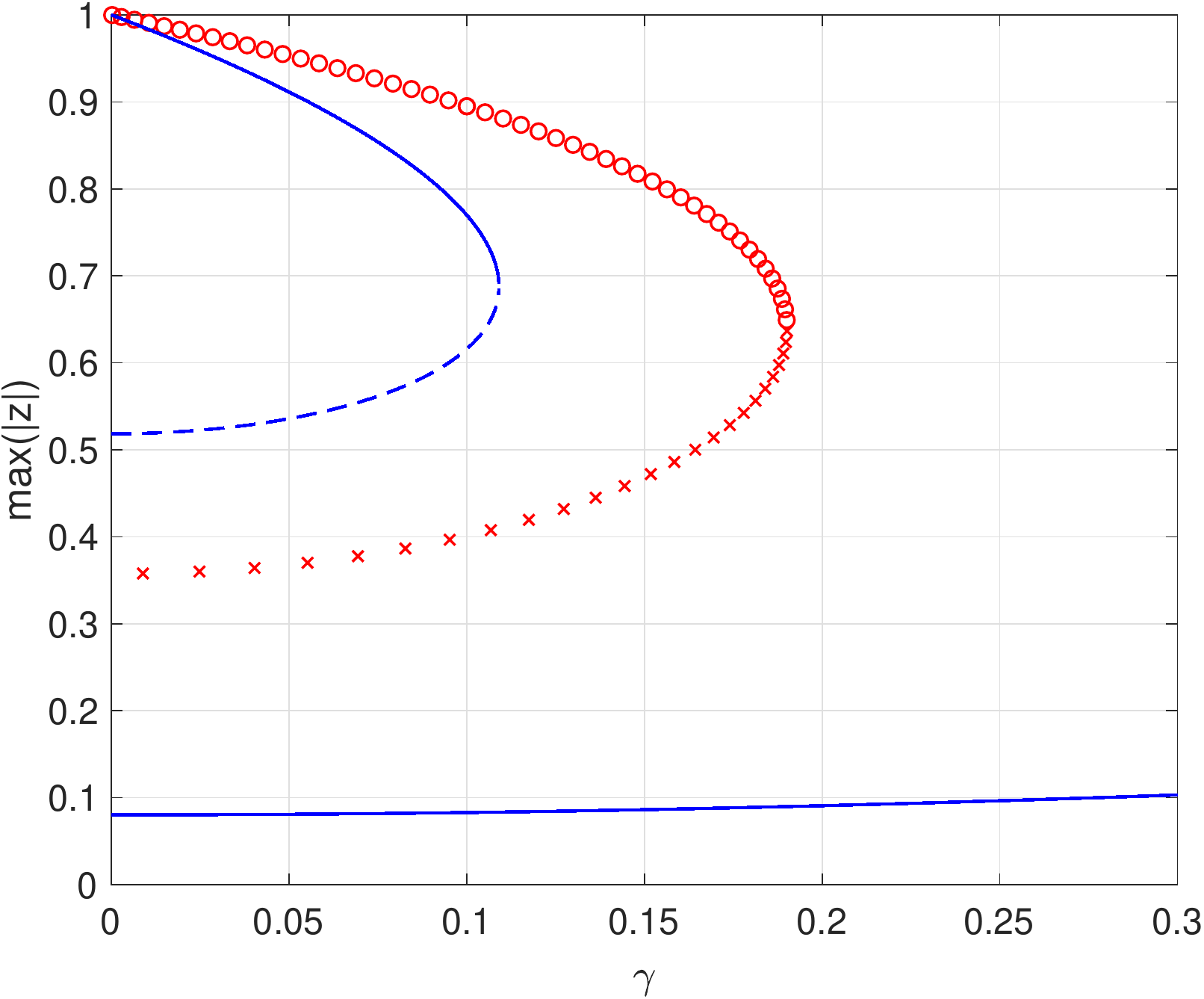}
\caption{Maximum over $x$ of $|z|$ for spatially-uniform states (blue curves; solid: stable,
dashed: unstable) and bump states (red symbols; circles: stable, crosses: unstable).
Other parameters: $\kappa=2$, $\eta_0=-0.4$, $n=2$. Note that the two types of solution
do not merge at $(\gamma,\max{|z|})=(0,1)$.}
\label{fig:sym}
\end{figure}

For the symmetric coupling kernel $K(x)$ and sufficiently small values~$\gamma$
equation~(\ref{Eq:NeuralField}) has three spatially uniform fixed points
$z(x) = \mathrm{const}.$ as well as a pair of spatially modulated bump states
$z(x) \ne \mathrm{const}$. Two of the spatially uniform states are stable
and differ in the level of synchrony between neurons.
They appear as spatially extended counterparts of partially synchronous rest (PSR)
and partially synchronous spiking (PSS) states reported in~\cite{lukbar13}.
More precisely, the PSR state is characterized by a higher level of synchrony ($|z|\approx 1$)
and mostly vanishing firing rates, see Fig.~\ref{fig:introduction}(a),
while in the PSS state the neurons fire almost asynchronously and uniformly in space, see Fig.~\ref{fig:introduction}(b).
The bump states, see Fig.~\ref{fig:introduction}(c), are spatially modulated states
comprising both active (firing) and quiescent spatial domains.
In equation~(\ref{Eq:NeuralField}) they appear as a pair of stable and unstable solutions.
As~$\gamma$ grows they move towards each other and eventually
collide and disappear in the saddle-node bifurcation for $\gamma\approx 0.19$:
see Fig.~\ref{fig:sym}. The same scenario occurs for the unstable spatially-uniform state
and the spatially-uniform PSR state.
In contrast, the PSS state exists as a separate stable branch for all values $\gamma$.

\subsection{Asymmetric coupling kernel $B\ne 0$}

\begin{figure}[t]
\includegraphics[scale=0.8]{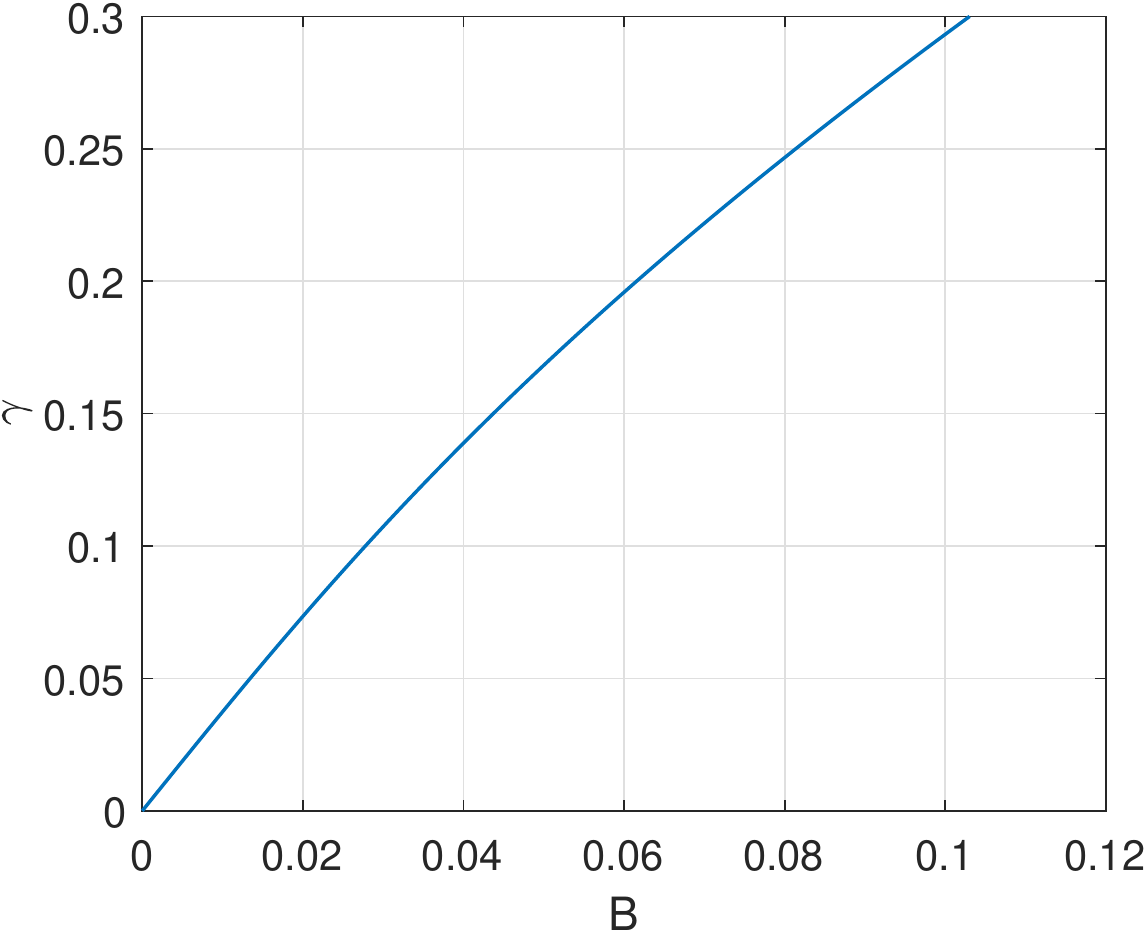}
\caption{Curve of Hopf bifurcation of spatially-uniform PSS state. This state is stable to the
left of this curve.
Other parameters: $\kappa=2$, $\eta_0=-0.4$, $n=2$.}
\label{fig:uni}
\end{figure}

Every symmetric solution to Eq.~(\ref{Eq:NeuralField}) existing for $B = 0$
remains a solution to this equation for an asymmetric coupling kernel too.
However, its stability properties may change as the asymmetry parameter~$B$ grows.
For example, for a fixed $\gamma$ there is a spatially-uniform PSS state, independent of $B$. 
However, it loses stability as $B$ is increased through a Hopf
bifurcation for which the eigenfunctions have spatial structure
i.e.~to a travelling wave, at points shown in Fig.~\ref{fig:uni}.

\begin{figure}[t]
\includegraphics[scale=0.6]{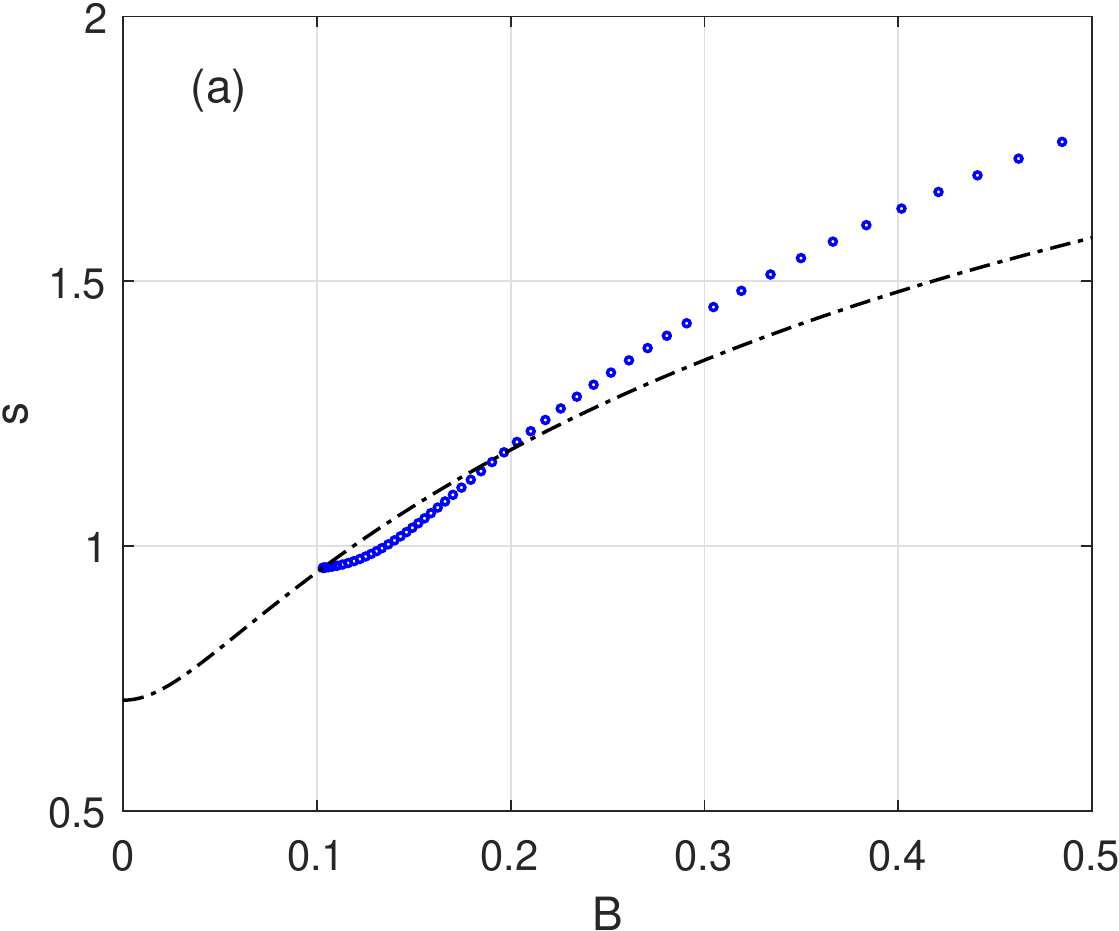}\hspace{10mm}
\includegraphics[scale=0.6]{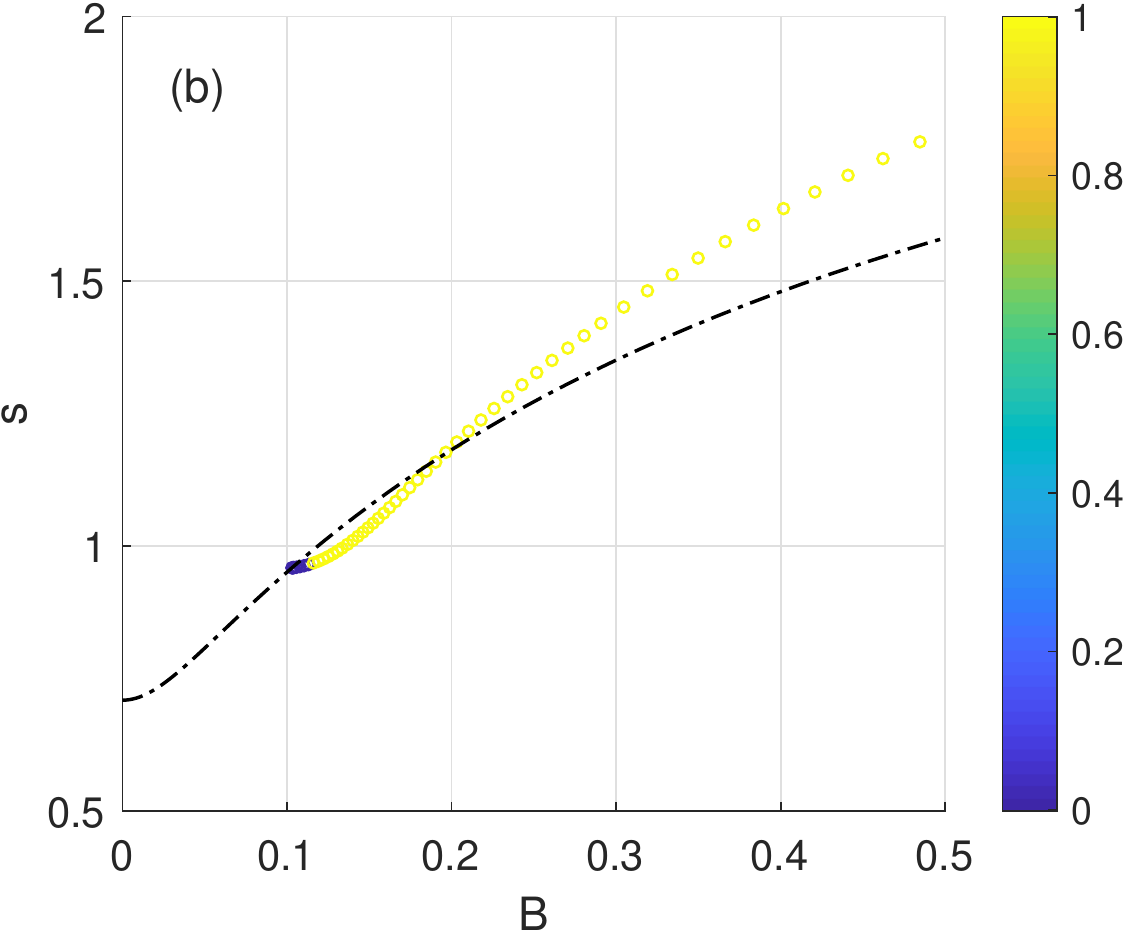}
\caption{Speed of travelling wave for $\gamma=0.3$.
(a) blue: stable. The dash-dotted line shows
the imaginary part of the Hopf eigenvalue of spatially uniform PSS state.
Panel (b) is identical to panel (a) but the twist of the travelling wave is indicated
using the colour bar on the right, rather than its stability.}
\label{fig:0_3a}
\end{figure}

\begin{figure}[t]
\includegraphics[scale=0.6]{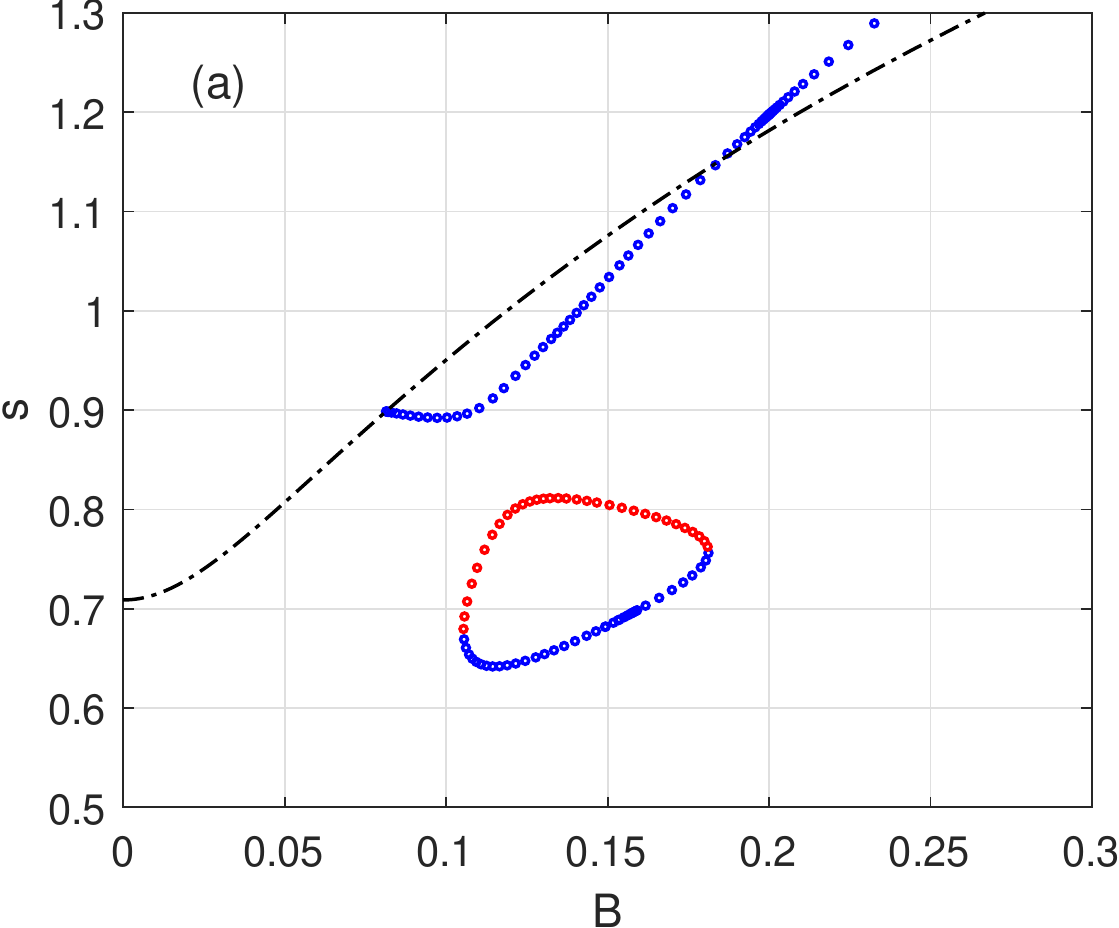}\hspace{10mm}
\includegraphics[scale=0.6]{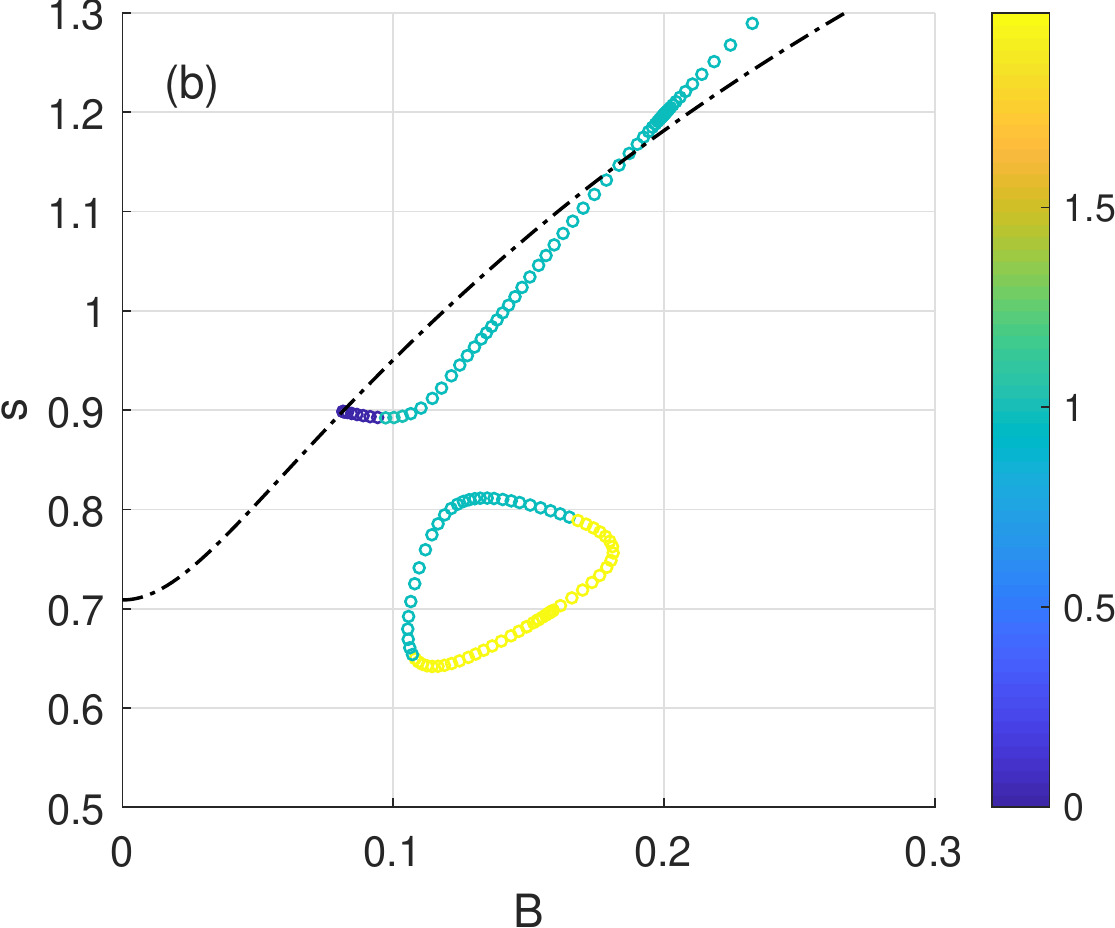}
\caption{Speed of travelling wave for $\gamma=0.25$.
(a) blue: stable, red: unstable. 
(b) the same graph but with twist indicated.}
\label{fig:0_25a}
\end{figure}

\begin{figure}[t]
\includegraphics[scale=0.6]{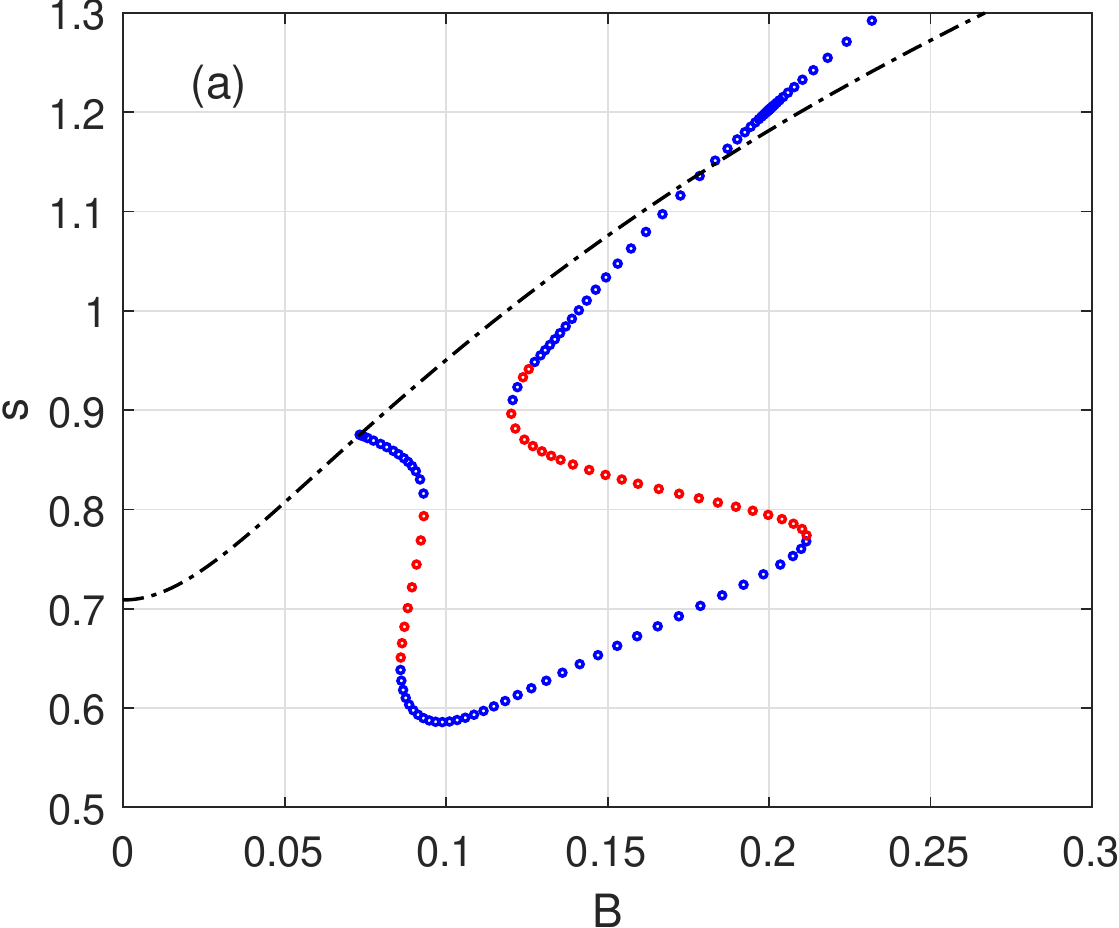}\hspace{10mm}
\includegraphics[scale=0.6]{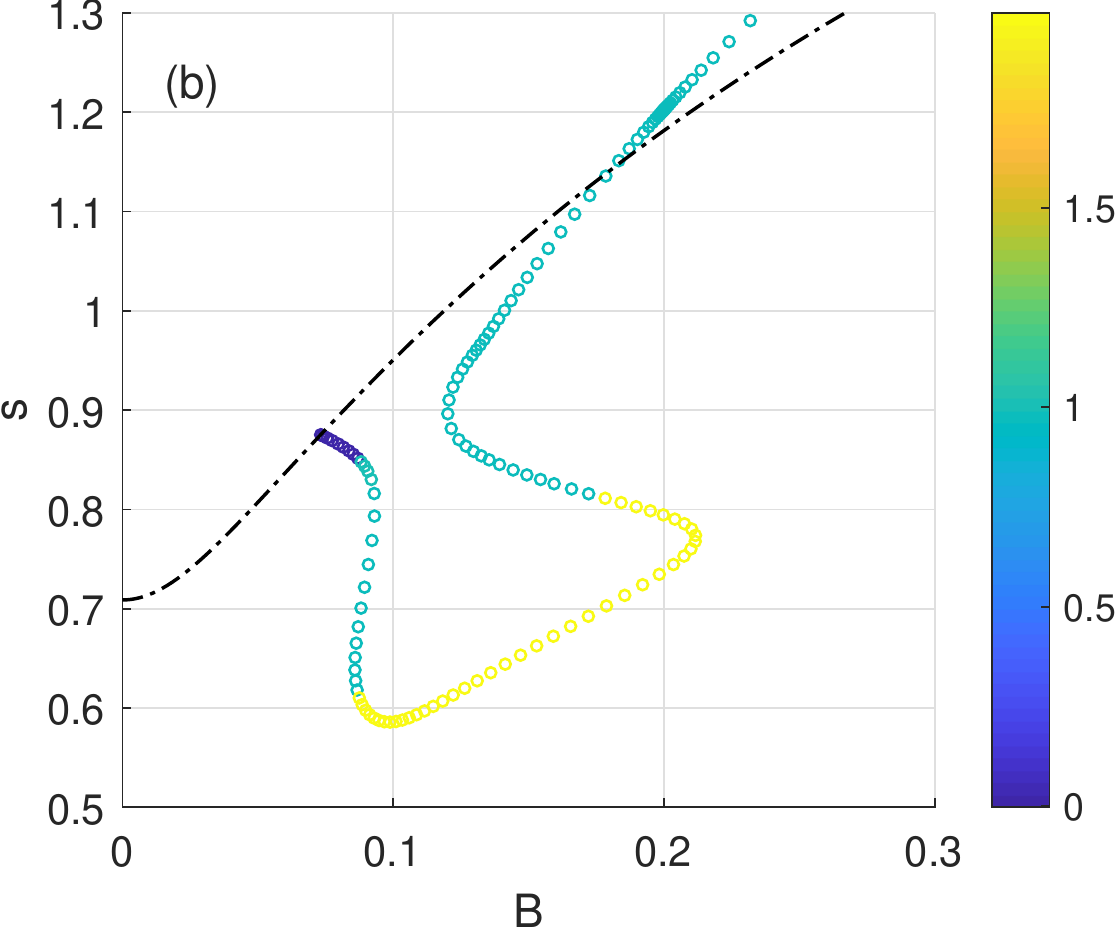}
\caption{Speed of travelling wave for $\gamma=0.23$.
(a) blue: stable, red: unstable. 
(b) the same graph but with twist indicated.}
\label{fig:0_23a}
\end{figure}

\begin{figure}[t]
\includegraphics[scale=0.6]{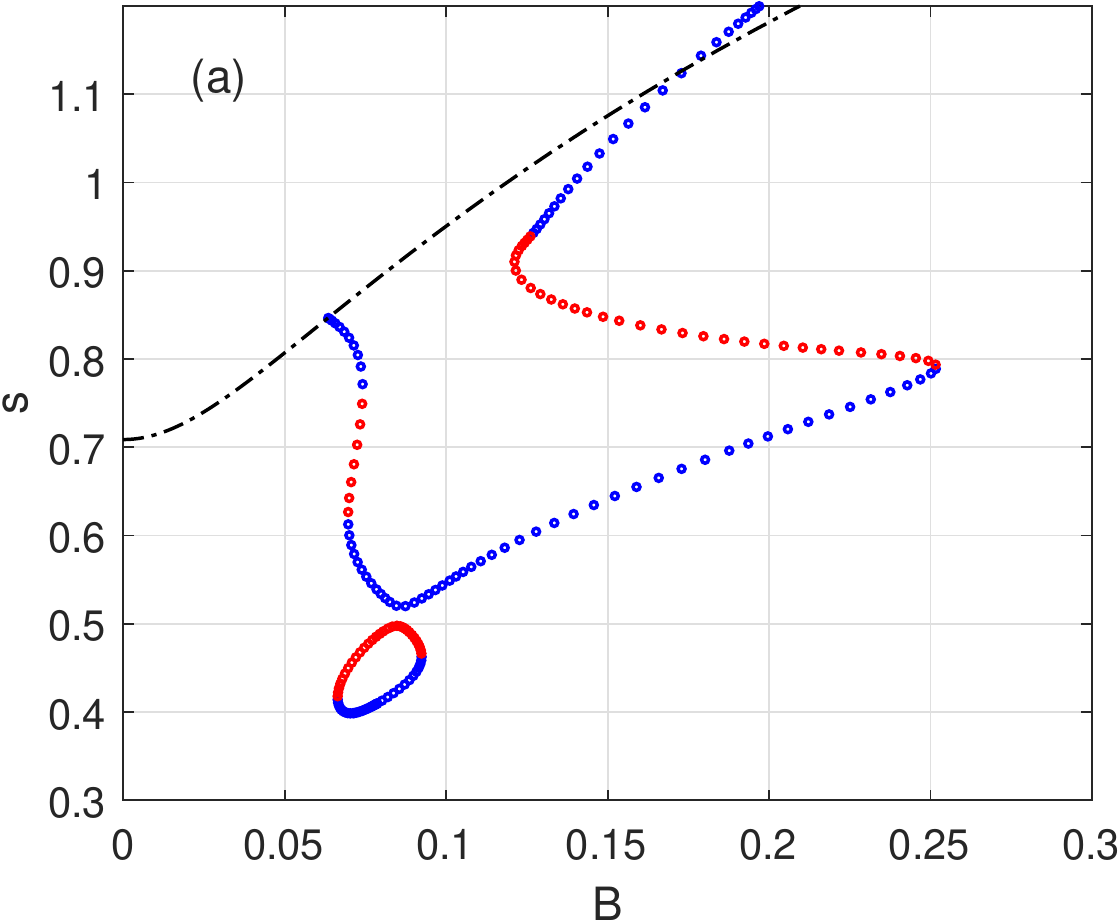}\hspace{10mm}
\includegraphics[scale=0.6]{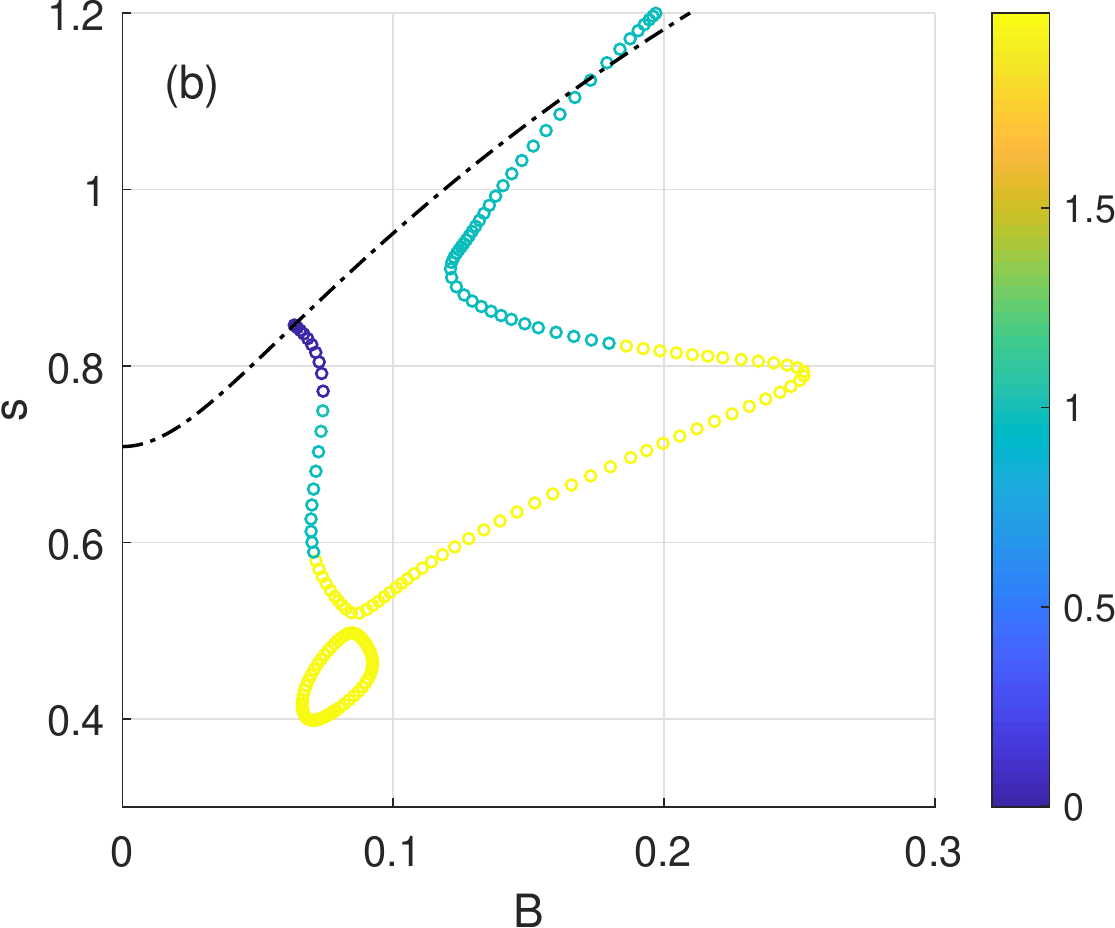}
\caption{Speed of travelling wave for $\gamma=0.205$.
(a) blue: stable, red: unstable. 
(b) the same graph but with twist indicated.}
\label{fig:0_205a}
\end{figure}

\begin{figure}[t]
\includegraphics[scale=0.6]{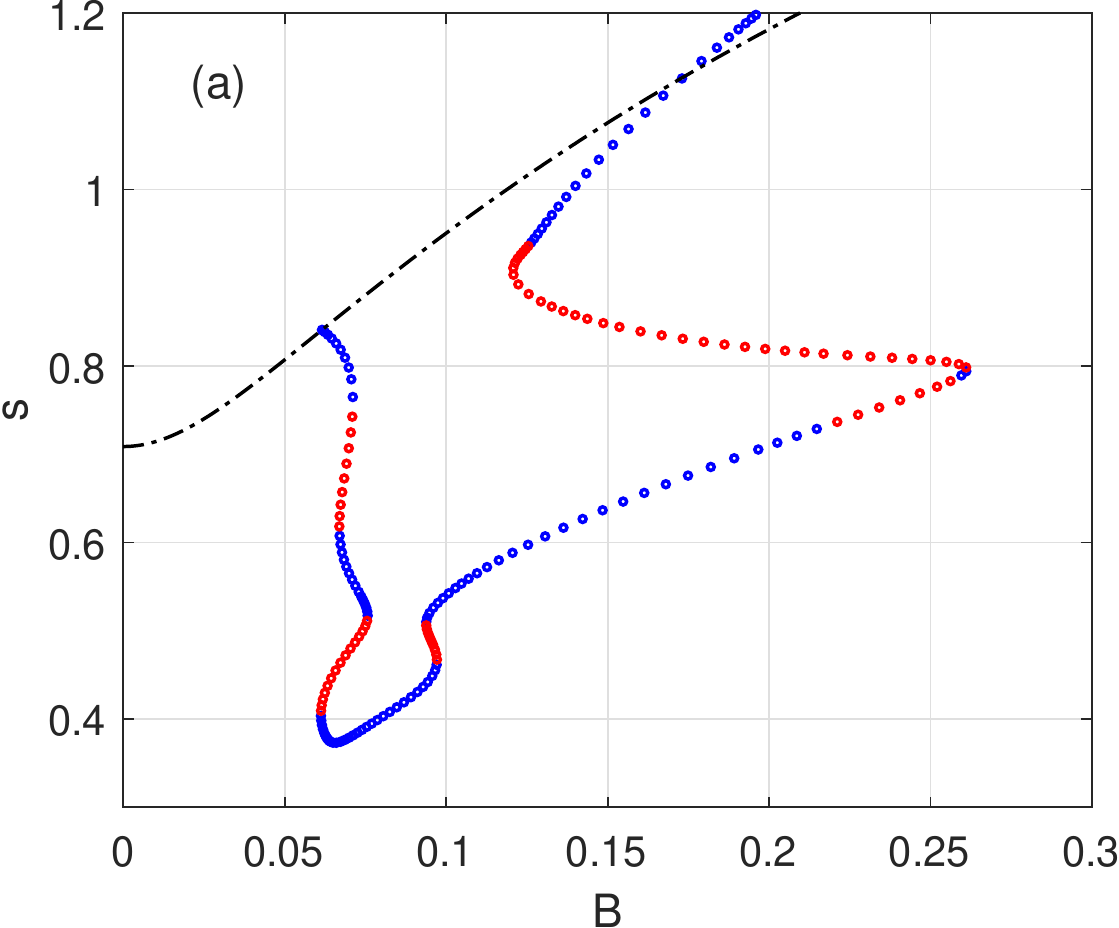}\hspace{10mm}
\includegraphics[scale=0.6]{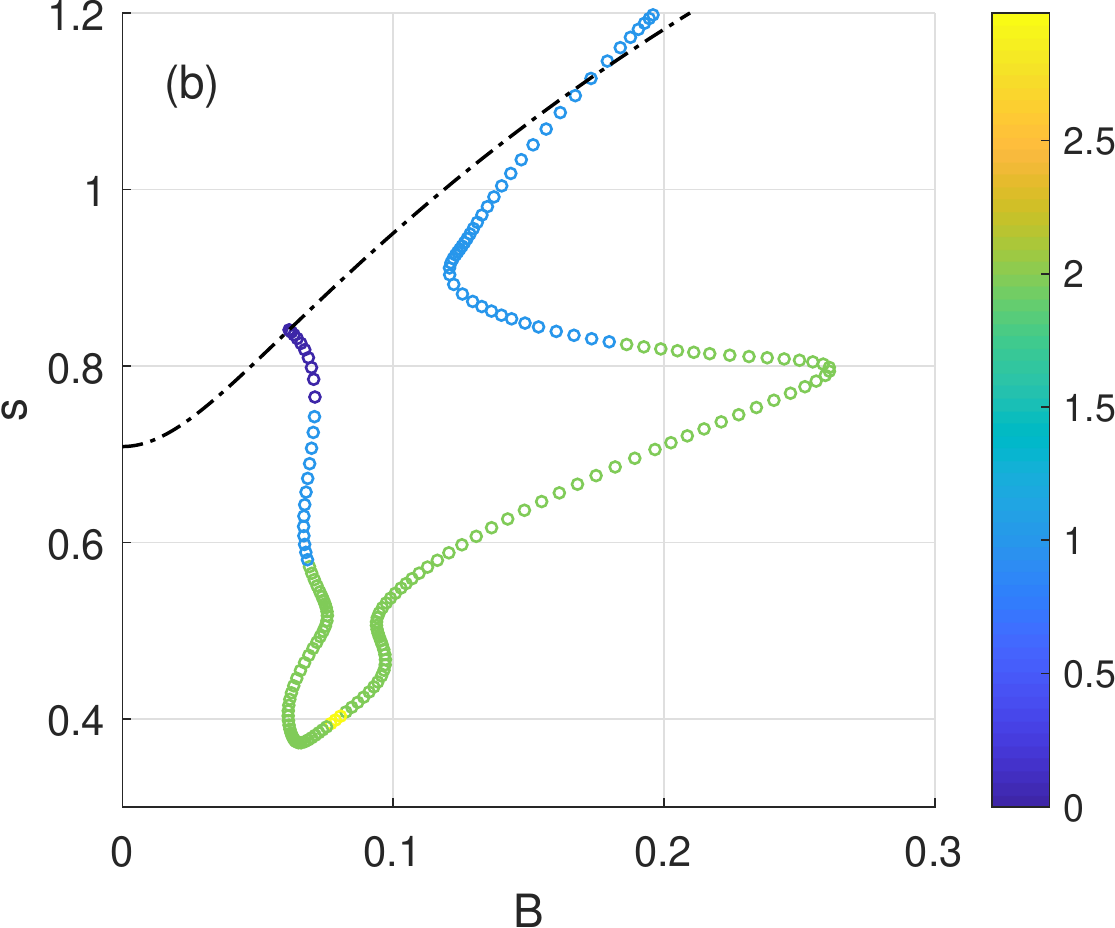}
\caption{Speed of travelling wave for $\gamma=0.2$.
(a) blue: stable, red: unstable. 
(b) the same graph but with twist indicated.}
\label{fig:0_2a}
\end{figure}

\begin{figure}[t]
\includegraphics[scale=0.6]{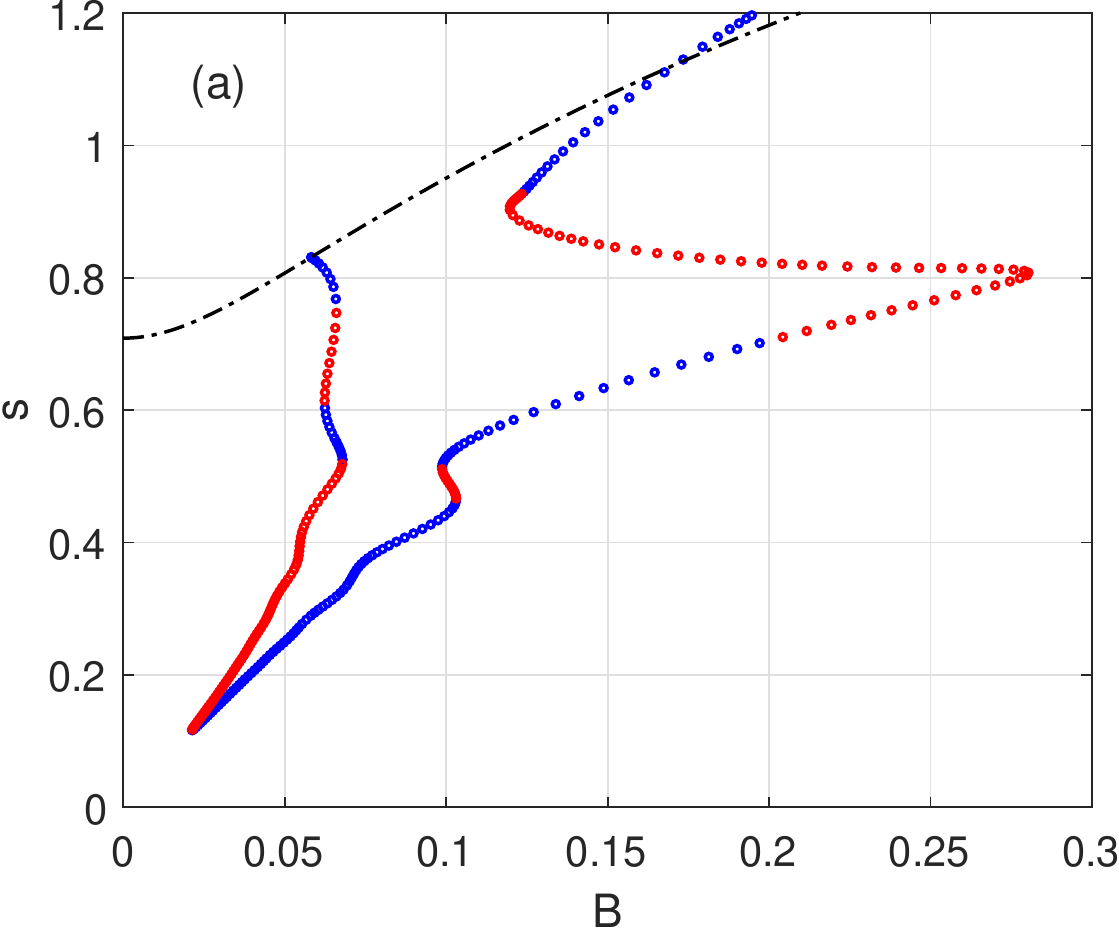}\hspace{10mm}
\includegraphics[scale=0.6]{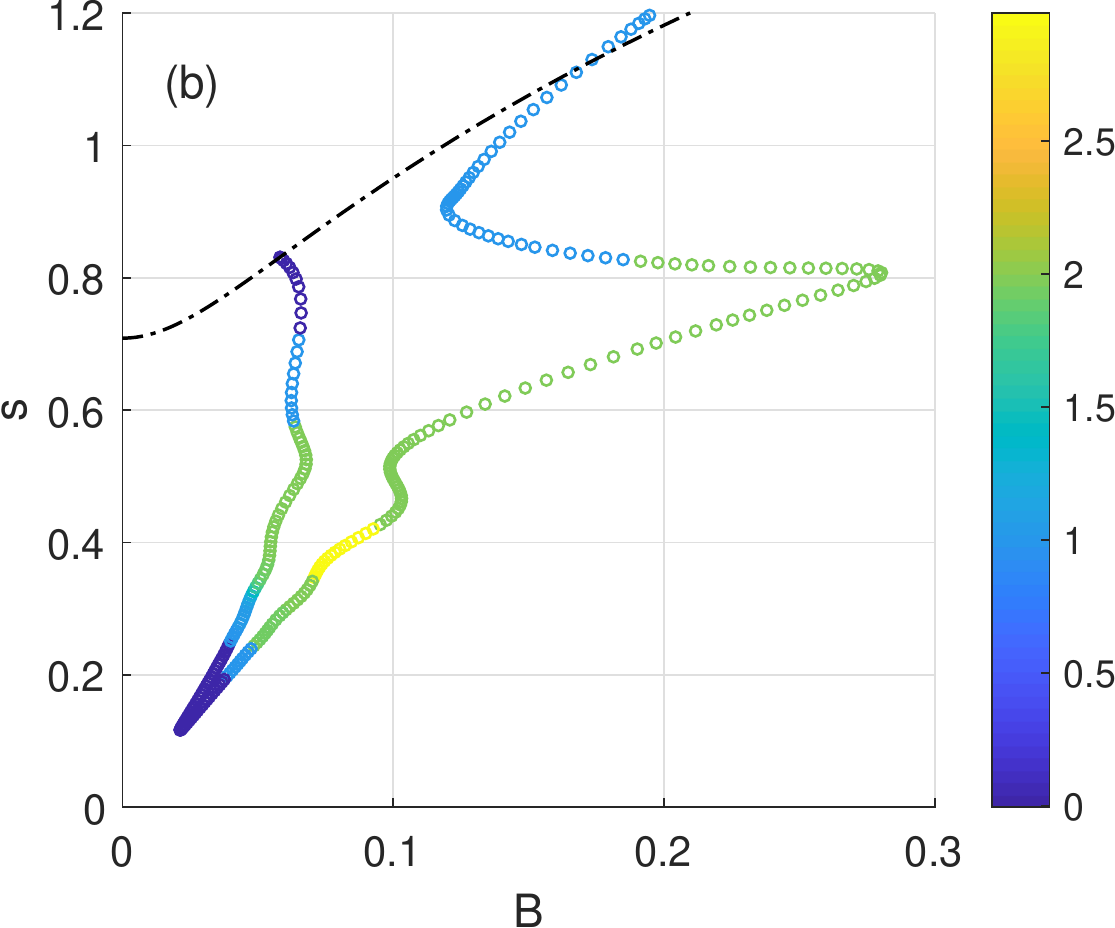}
\caption{Speed of travelling wave for $\gamma=0.191$.
(a) blue: stable, red: unstable. 
(b) the same graph but with twist indicated.}
\label{fig:0_191a}
\end{figure}

\begin{figure}[t]
\includegraphics[scale=0.6]{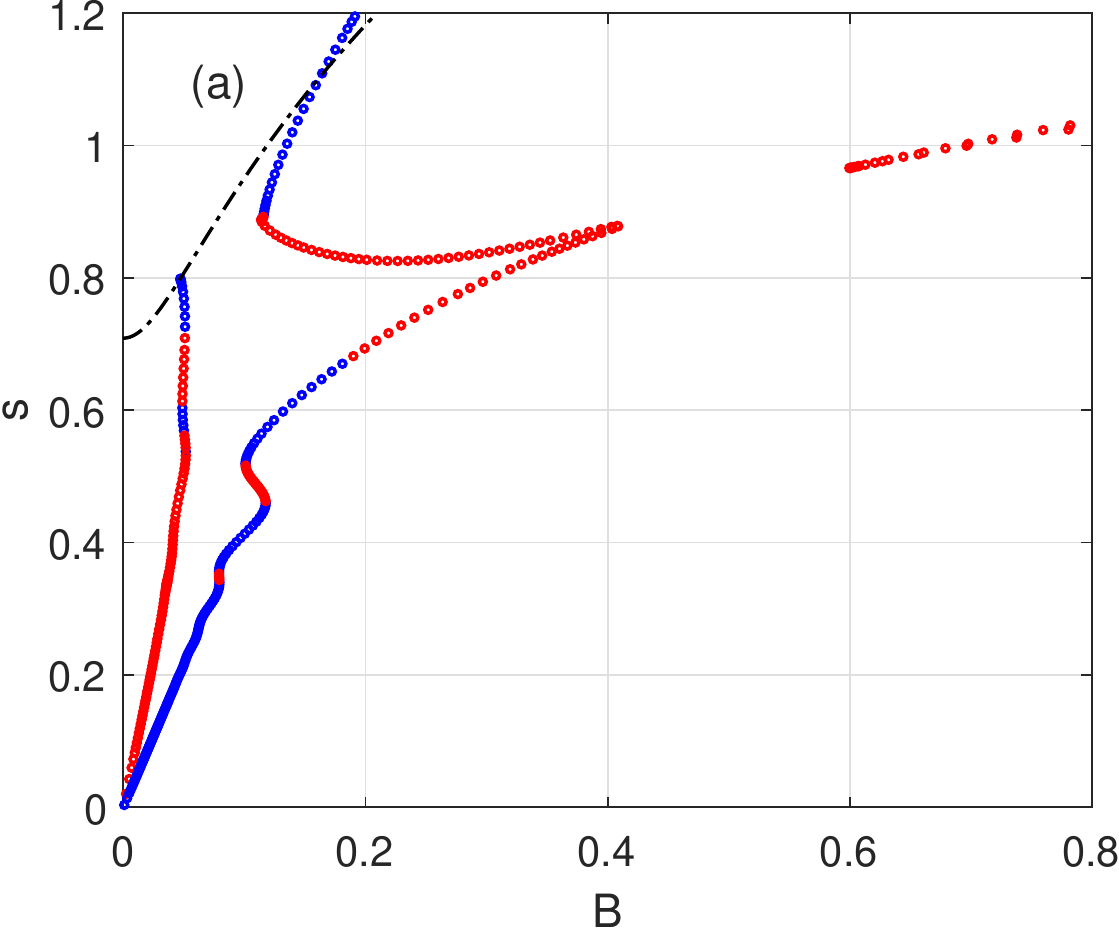}\hspace{10mm}
\includegraphics[scale=0.6]{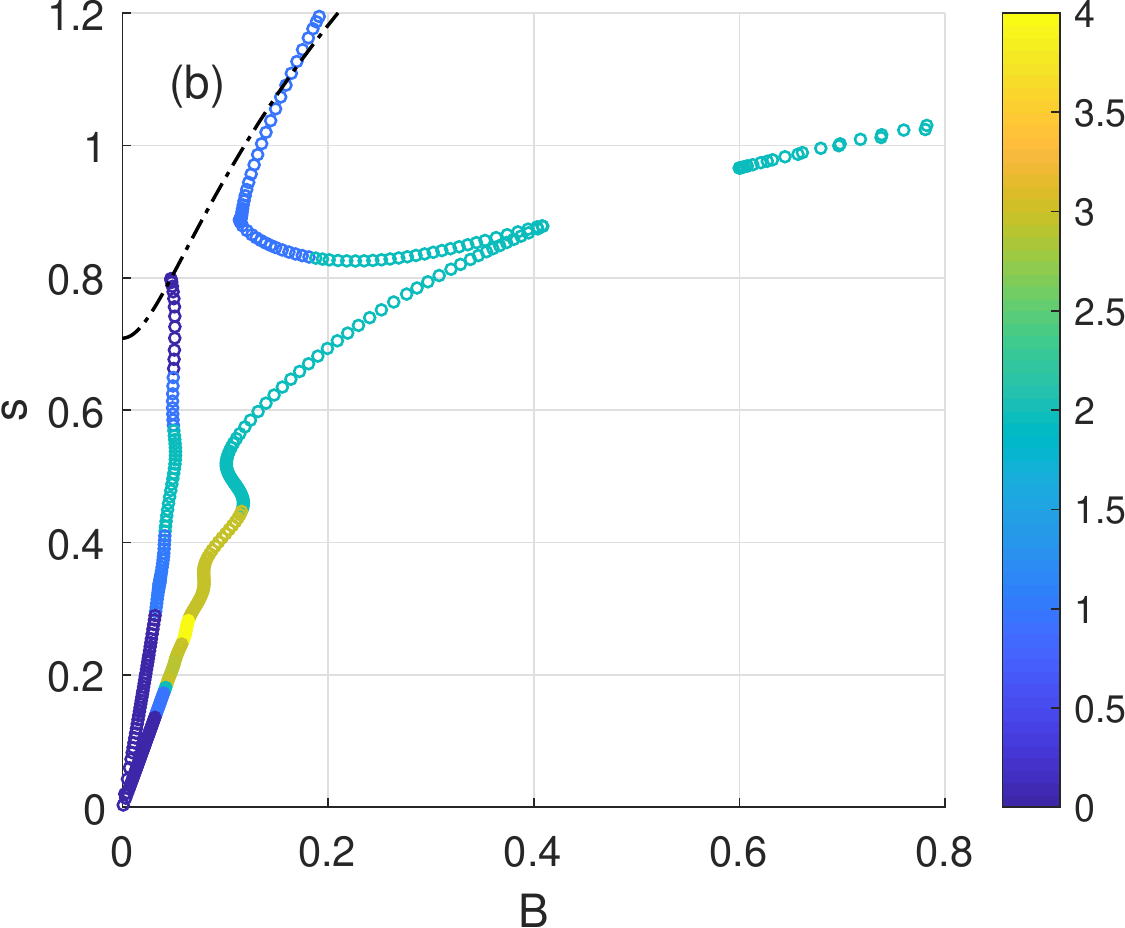}
\caption{Speed of travelling wave for $\gamma=0.16$.
(a)  blue: stable, red: unstable. There are two branches
with similar speeds for large~$B$. Two branches leave the origin.
(b) the same graph but with twist indicated.}
\label{fig:0_16a}
\end{figure}

\begin{figure}[t]
\includegraphics[scale=0.6]{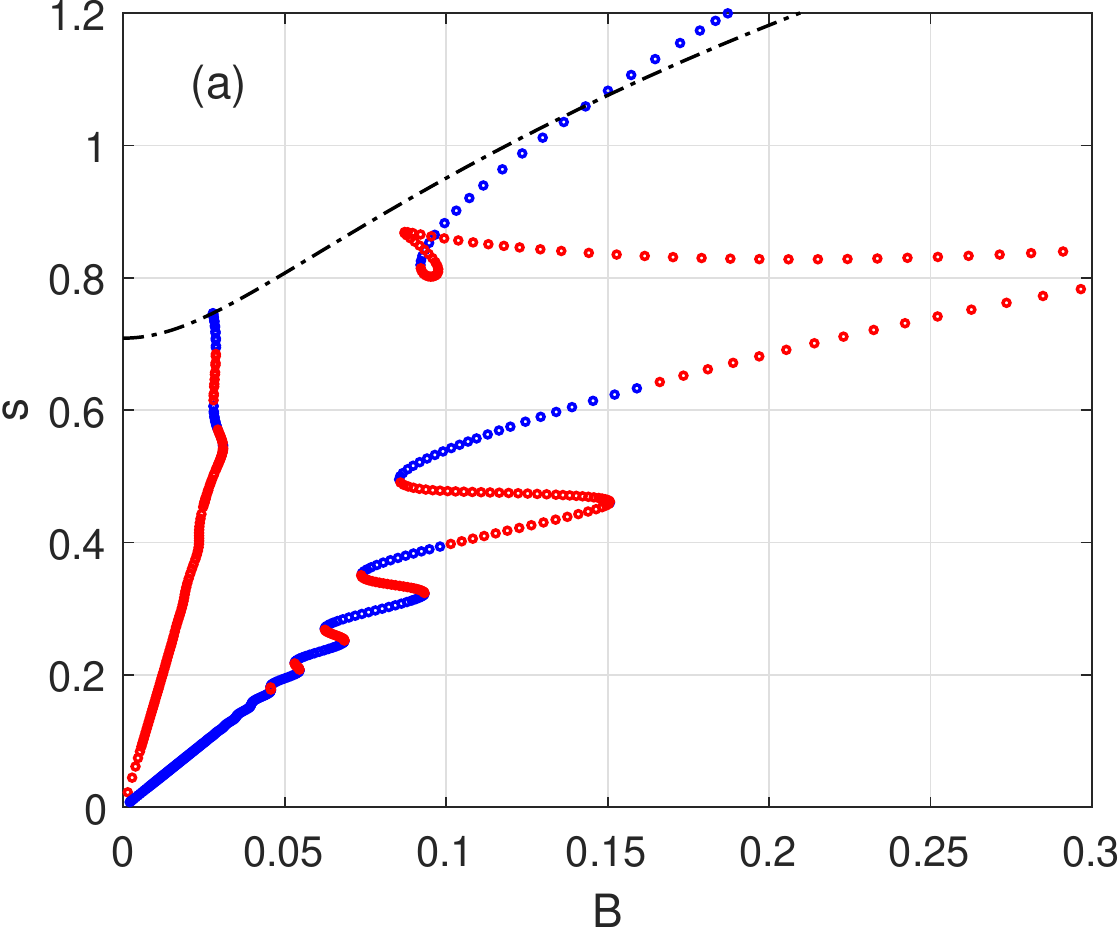}\hspace{10mm}
\includegraphics[scale=0.6]{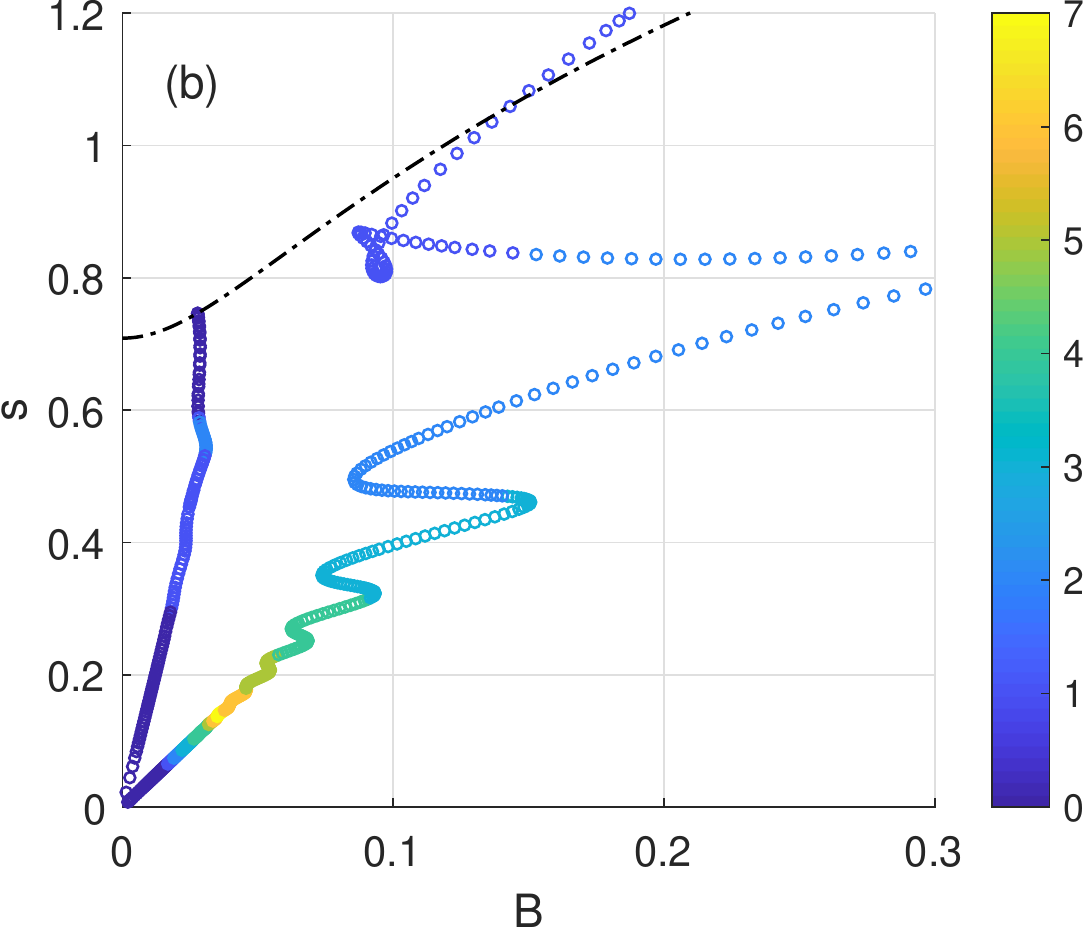}
\caption{Speed of travelling wave for $\gamma=0.1$.
(a) blue: stable, red: unstable. 
(b) the same graph but with twist indicated.}
\label{fig:0_1a}
\end{figure}

\begin{figure}[t]
\includegraphics[scale=0.8]{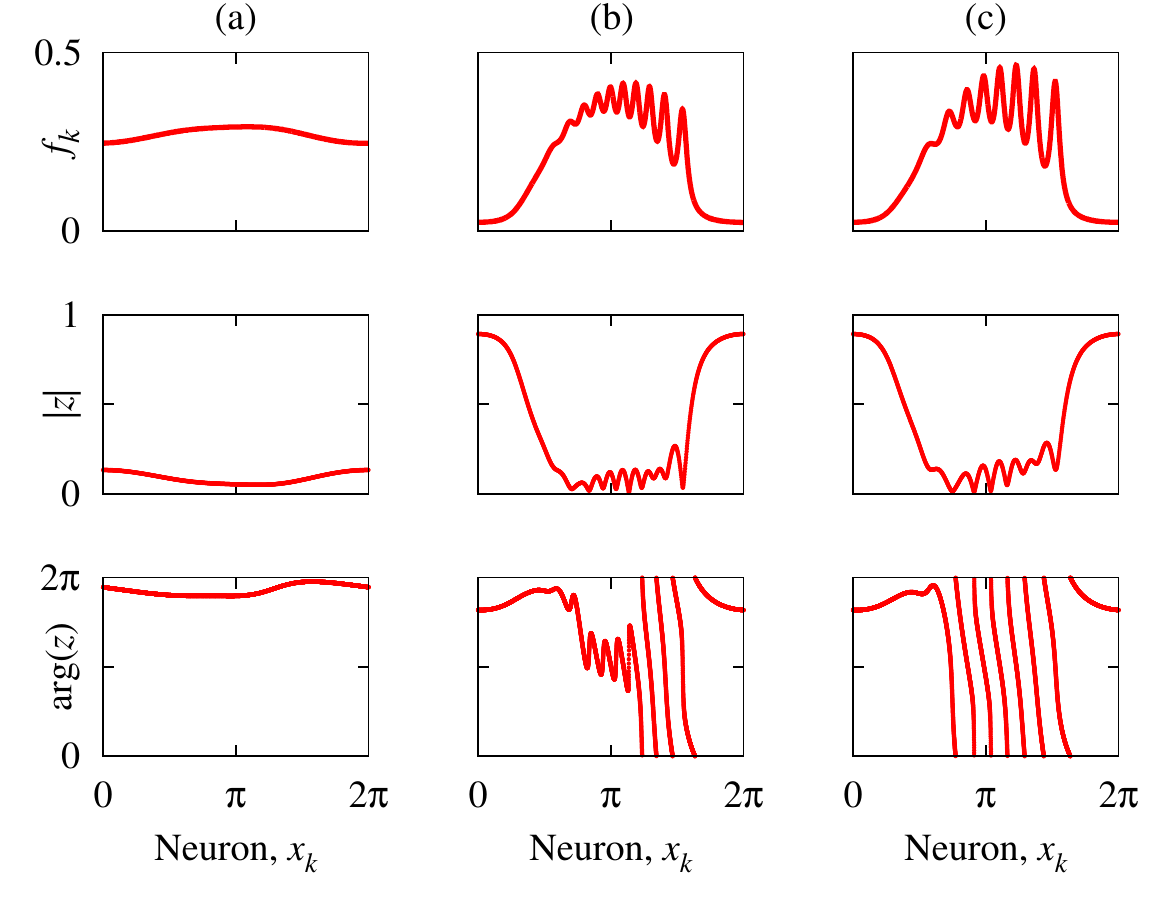}
\caption{
Travelling wave solutions to the neural field Eq.~(\ref{Eq:NeuralField}).
Columns~(a) and~(b) show the fastest and the slowest solutions, respectively, 
coexisting stably for $B = 0.0285$;
column~(c) shows a stable solution for $B = 0.036$.
All three travelling waves are moving to the right.
Other parameters: $\kappa = 2$, $\eta_0 = -0.4$, $n = 2$ and $\gamma=0.1$.
Top row: instantaneous frequency $f$ (see~(\ref{Formula:f})). Middle row: $|z|$. Bottom row: arg($z$).}
\label{fig:examp}
\end{figure}

\begin{figure}[t]
\includegraphics[scale=0.6]{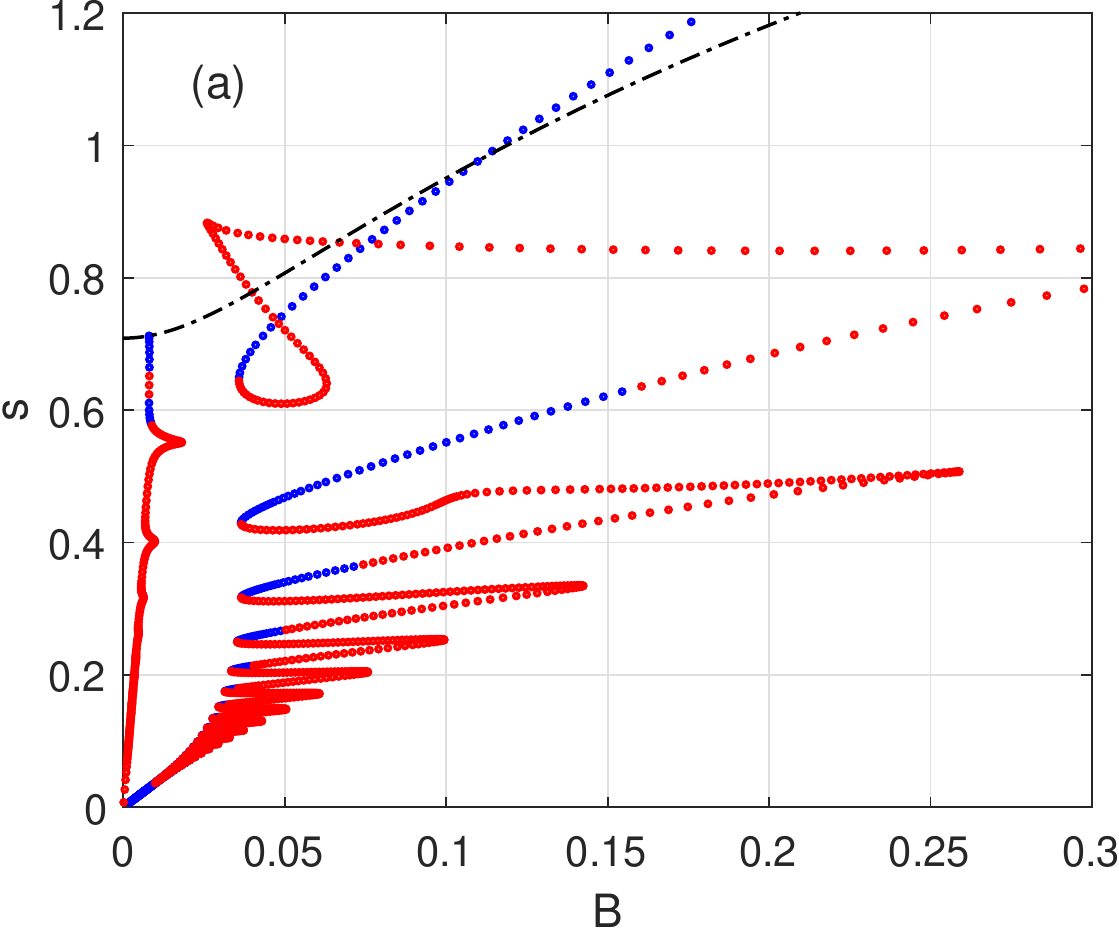}\hspace{10mm}
\includegraphics[scale=0.6]{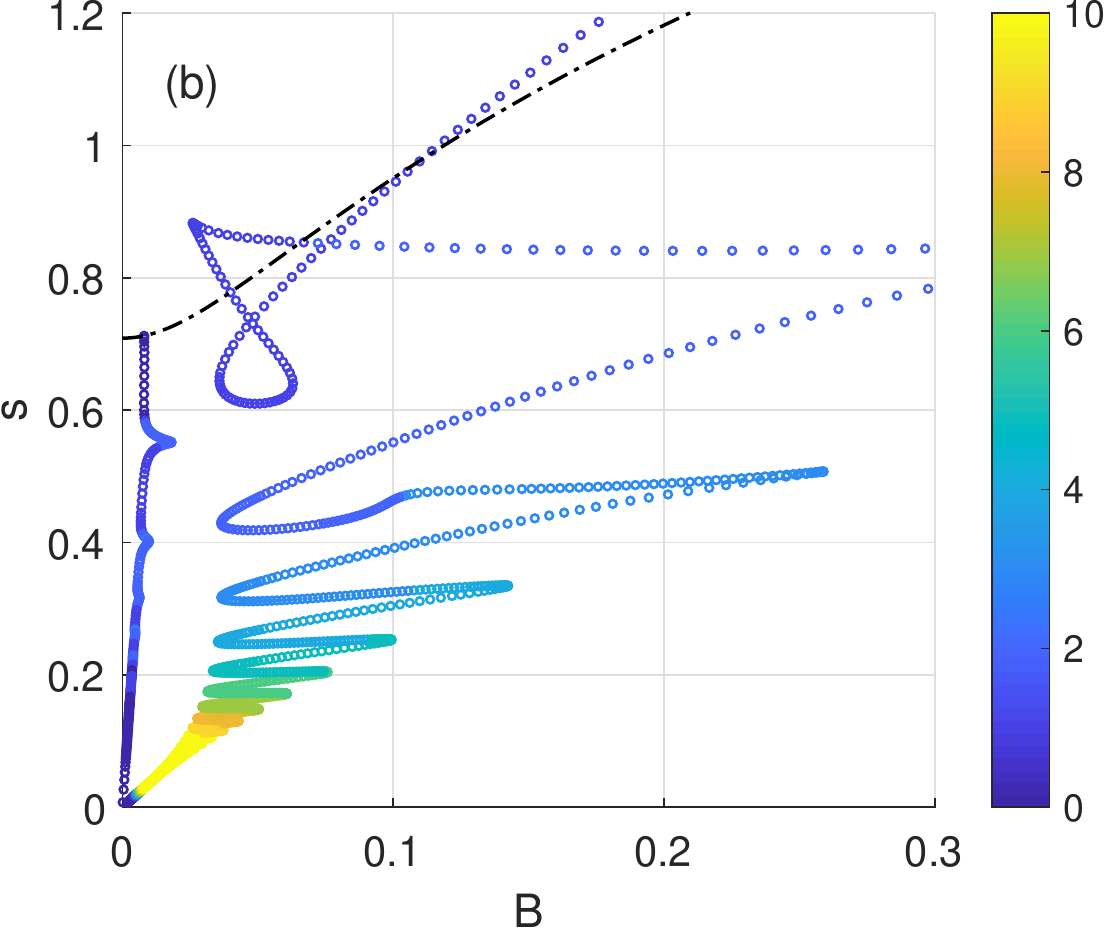}
\caption{Speed of travelling wave for $\gamma=0.03$.
(a) blue: stable, red: unstable. 
(b) the same graph but with twist indicated.}
\label{fig:0_03a}
\end{figure}

In Figs.~\ref{fig:0_3a}--\ref{fig:0_03a} we plot the imaginary part~$\omega$
of the Hopf eigenvalue~$\lambda_\mathrm{H}$ as a function of $B$ (dash-dotted curve).
At the point of bifurcation the branch of travelling waves
bifurcates from this curve and the speed of the travelling wave is $\omega$.
For all of these figures we use $\kappa=2$, $\eta_0=-0.4$, $n=2$.

We now describe the series of transitions that occur to the bump states as $\gamma$ is decreased.
For $\gamma=0.3$ there is a single stable branch of travelling solutions which start out
as twist-0 but become twist-1 as $B$ is increased (see Fig.~\ref{fig:0_3a}). 
Between $\gamma=0.3$ and $\gamma=0.25$
another family of solutions is created which form an isola (see Fig.~\ref{fig:0_25a}). 
Decreasing $\gamma$ to 0.23 these two
branches merge, resulting in one branch of solutions with twists 0,1 and 2 (see Fig.~\ref{fig:0_23a}).
At $\gamma=0.205$ another isola has been created (Fig.~\ref{fig:0_205a}), 
and this merges with the main branch
as $\gamma$ is decreased to 0.2, and the first twist-3 solution appears (see Fig.~\ref{fig:0_2a}).
For $\gamma=0.25$ the only changes of stability are due to saddle-node bifurcations
but for $\gamma\leq 0.23$ Hopf bifurcations also occur, so branches do not simply
alternate in stability as $B$ increases and decreases.

At $\gamma=0.191$ the main branch of solutions nearly reaches the origin (Fig.~\ref{fig:0_191a})
and for $\gamma\leq 0.19$ there are two branches which leave the origin,
one stable one with lower speed, and an unstable one with higher speed.
Notice that this value $\gamma$ corresponds to that of the saddle-node bifurcation
of stable and unstable bumps observed in the symmetric coupling kernel case.
From Fig.~\ref{fig:sym} we know that for $B=0$ and $\gamma\leq 0.19$ there are two stationary 
bumps ($s=0$),
one stable and one unstable.

As $\gamma$ is decreased further an unstable branch appears from large $B$ (Fig.~\ref{fig:0_16a})
and merges with the main branch. For $\gamma\leq 0.16$ we see that
on the stable branch leaving the origin the twist increases
and then decreases as the speed increases. 
The main curve also develops a ``knot'' at approximately $B=0.1$ and speed $0.85$
(see Fig.~\ref{fig:0_1a}).
As an example of the types of solution found we show several stable solutions
for $\gamma=0.1$ in Fig.~\ref{fig:examp};
other examples can be also found in Fig.~\ref{fig:moving_bumps}.
In particular, two stable solutions coexisting for $B = 0.0285$ are shown in Fig.~\ref{fig:examp}(a),(b)
(the third solution coexisting stably for the same parameters is omitted).
The solution on the upper branch (Fig.~\ref{fig:examp}(a)) looks rather as a spatially modulated
version of the PSS state, while the lower branch solution (Fig.~\ref{fig:examp}(b))
is recognisably a ``bump'' although with many oscillations in instantaneous frequency
in the active part of the bump. Obviously, the latter has twist 4.
As $B$ grows the twist value grows too and achieves its maximum value of 7 for $B\approx 0.036$.
The corresponding twist-7 solution is shown in Fig.~\ref{fig:examp}(c).
For further growth of~$B$ the twist value of the bump starts to decrease, Fig.~\ref{fig:0_1a}(b),
and the speed-asymmetry curve is not a simple function of $B$.
Notice that the twist-3, twist-2 and twist-1 solutions found for larger asymmetry, Fig.~\ref{fig:moving_bumps},
could perhaps be better thought of as ``3-shot, 2-shot and 1-shot waves''
in which neurons at a fixed position have several bursts of activity during one complete pass of the wave.
Remarkably, the behavior of the lower branch in Fig.~\ref{fig:0_1a} resembles
the speed-asymmetry diagram computed for travelling chimera states
in asymmetrically coupled phase oscillators~\cite[Fig.~4]{ome20}.
However, the complete bifurcation scenario of moving bumps in general is more complicated,
which follows apparently from the complexity of the neural field equation~(\ref{Eq:NeuralField})
in comparison with the mean field equation analyzed in~\cite{ome20}.

\begin{figure}[t]
\includegraphics[scale=0.6]{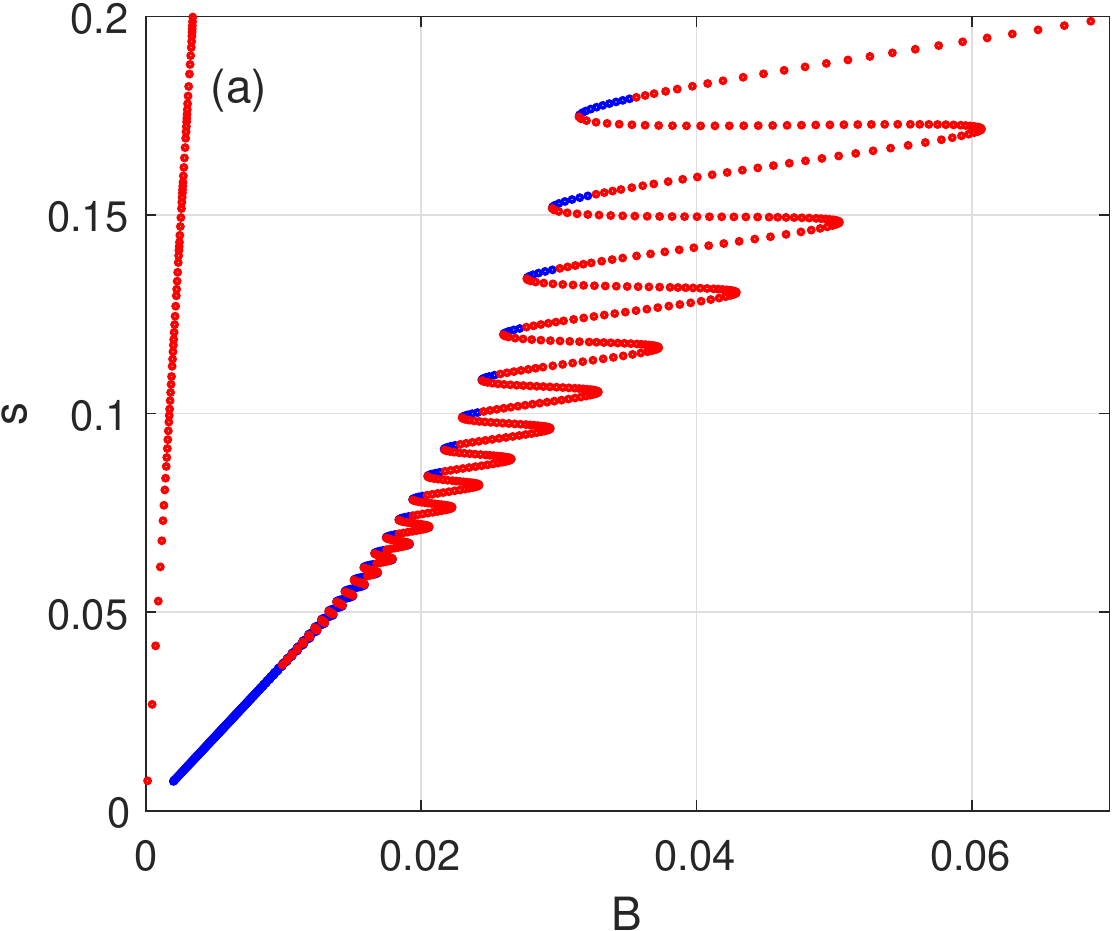}\hspace{10mm}
\includegraphics[scale=0.6]{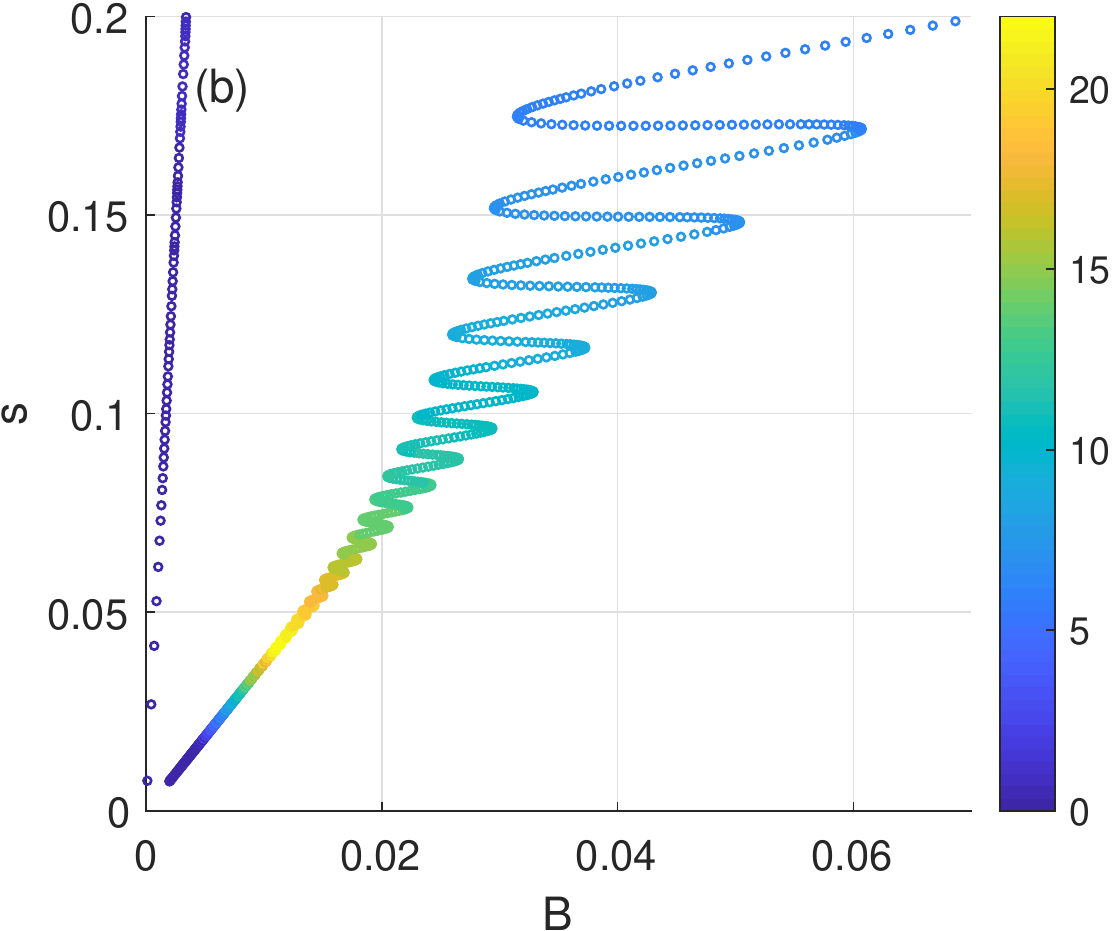}
\caption{A zoom of Fig.~\ref{fig:0_03a}.
Speed of travelling wave for $\gamma=0.03$.
(a) blue: stable, red: unstable. 
(b) the same graph but with twist indicated.
(Both branches pass through the origin.)}
\label{fig:0_03az}
\end{figure}

\begin{figure}[t]
\includegraphics[scale=0.6]{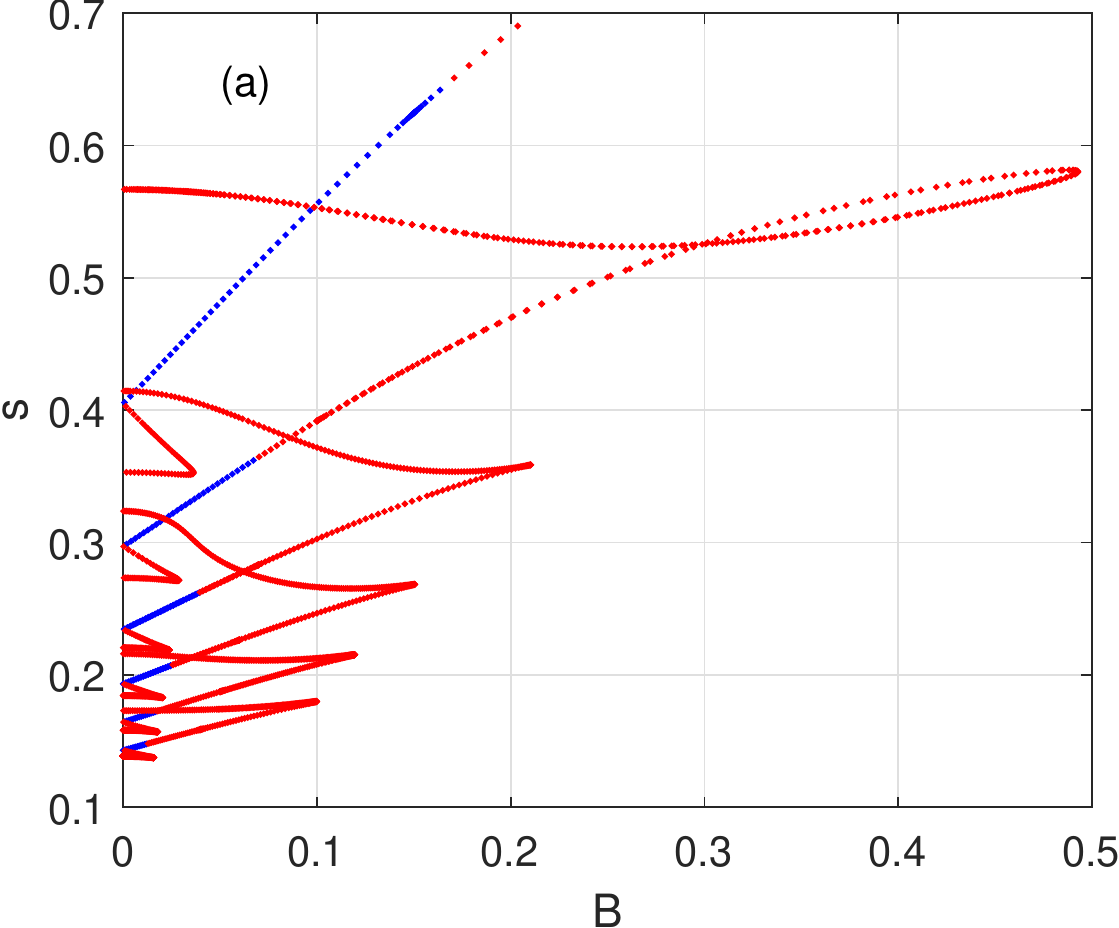}\hspace{10mm}
\includegraphics[scale=0.6]{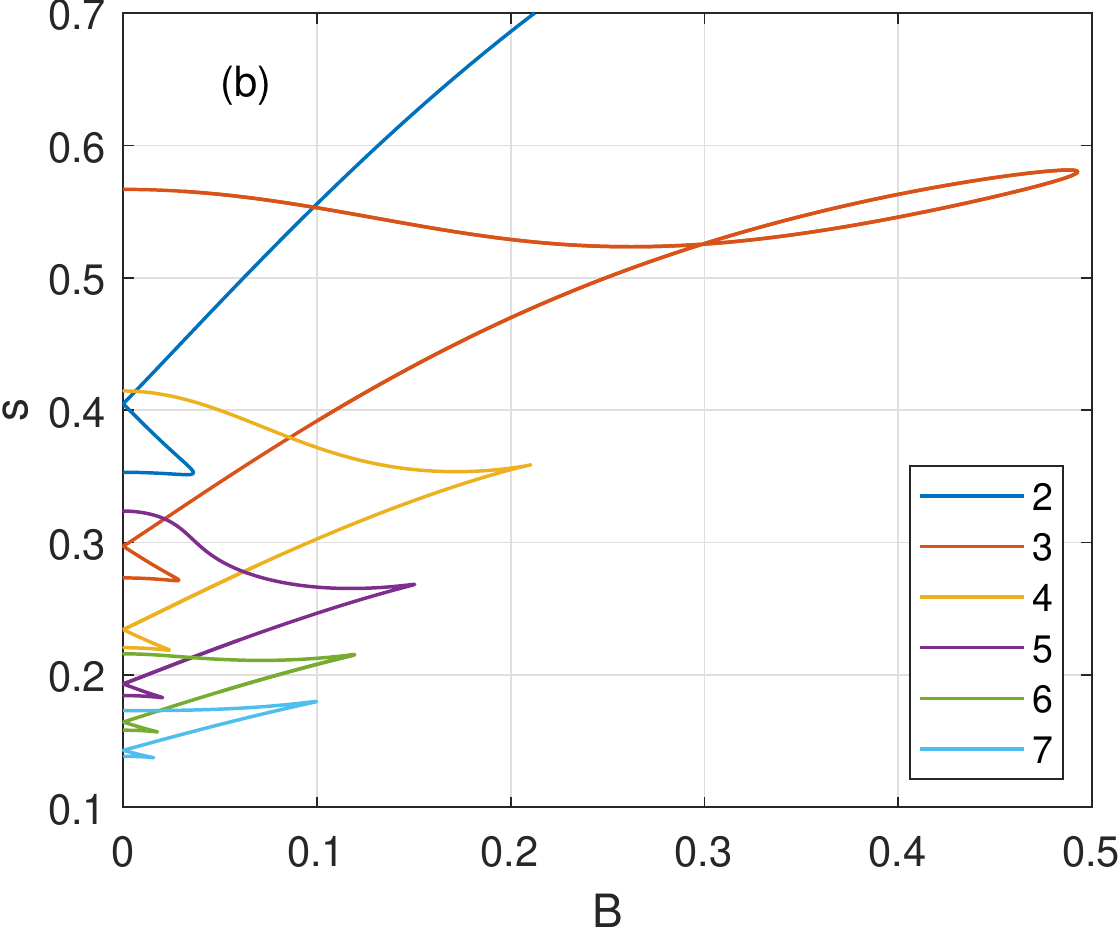}
\caption{Speed of travelling wave for $\gamma=0$.
(a) blue: stable, red: unstable. All instabilities are Hopf bifurcations.
(b) the same graph but with twist indicated.}
\label{fig:gam0}
\end{figure}

As $\gamma$ is decreased further the stable branch emanating from the origin develops
more and more oscillations (Figs.~\ref{fig:0_03a}--\ref{fig:0_03az}) and the maximum twist
value on this branch increases. For $\gamma=0$ the branches with different twists
are disconnected as shown in Fig.~\ref{fig:gam0} (only waves with twists 2-7 are shown).
These waves have $|z|=1$. 
It seems that a branch with twist $k$ is stable in an interval $B\in[0,B_k]$
where, numerically, $B_k\sim k^{-2.07}$.
The existence of these solutions is in contrast to the results
in~\cite{ome19} where it was found that travelling chimera states have no well-defined
limiting behaviour for $\gamma\to 0$.

%While it is not clear what happens to this branch
%as $\gamma\rightarrow 0$, other solutions exist and are well-defined for $B\neq 0$ and
%$\gamma=0$ (not shown). 

\begin{figure}[t]
\includegraphics[scale=0.8]{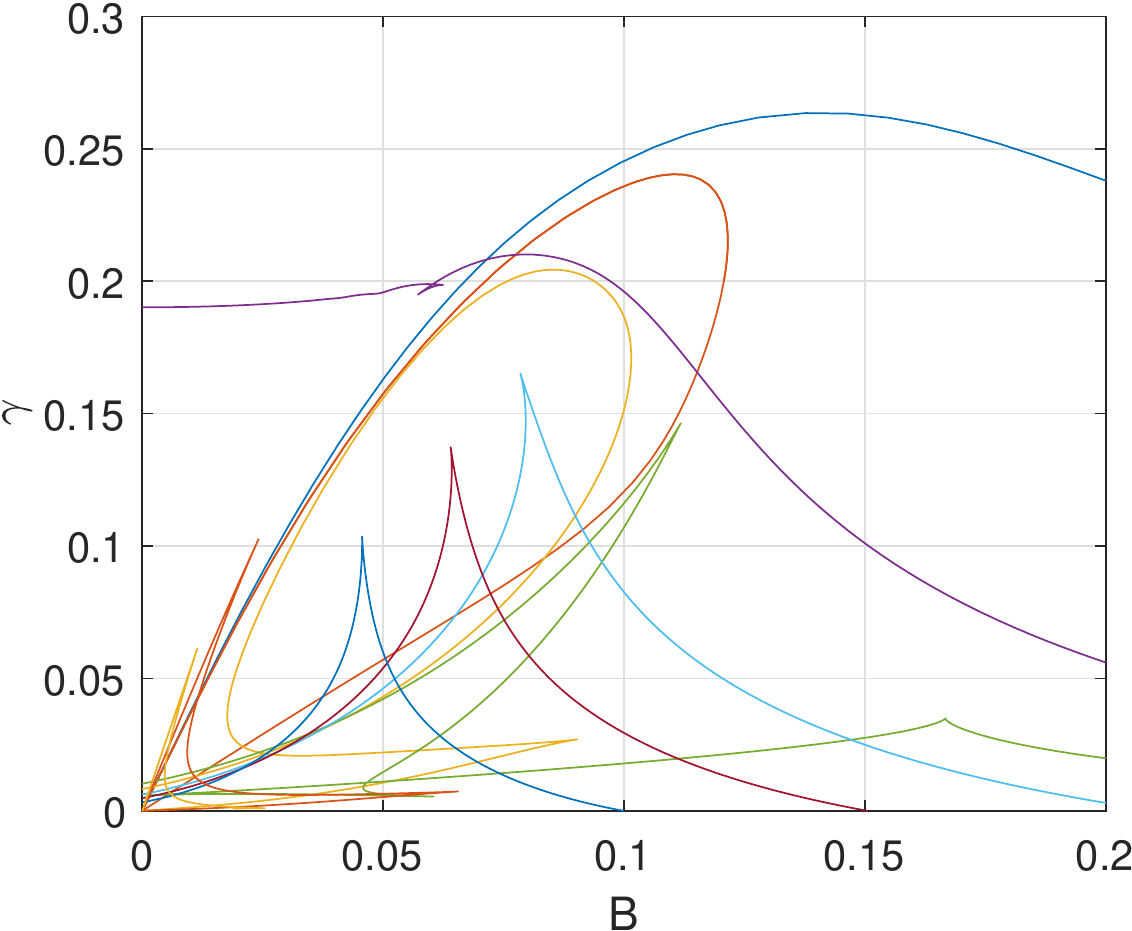}
\caption{Some of the saddle-node bifurcation curves corresponding to creation/destruction
of solutions shown in Figs.~\ref{fig:0_3a}--\ref{fig:0_1a} and~\ref{fig:0_03a}--\ref{fig:gam0}.
For example, at $\gamma=0.25$ two bifurcations occur at $B\approx 0.11$ and $B\approx 0.18$
(blue curve)
and these correspond to the edges of the isola seen in Fig.~\ref{fig:0_25a}. Similarly,
at $\gamma=0.23$ three bifurcations are seen (another occurs for $B > 0.2$) which
correspond to the four seen in Fig.~\ref{fig:0_23a}.}
\label{fig:sn1}
\end{figure}

The results are summarised in Fig.~\ref{fig:sn1}, where many of the
saddle-node bifurcations shown in Figs.~\ref{fig:0_3a}--\ref{fig:0_1a} 
and~\ref{fig:0_03a}--\ref{fig:gam0} are continued
as both $\gamma$ and $B$ are varied. There appears to be some self-similarity in the
arrangement of saddle-node bifurcations, but we have not investigated this further.

%\begin{figure}[t]
%\includegraphics[scale=0.8]{sn2}
%\caption{Some of the saddle-node bifurcation curves corresponding to creation/destruction
%of solutions shown in Figs.~\ref{fig:0_3a}-\ref{fig:0_03a}.}
%\label{fig:sn2}
%\end{figure}

\section{Discussion}
\label{sec:disc}
We numerically investigated moving bump solutions in the continuum equations describing
an infinite network of theta neurons nonlocally coupled through the asymmetric kernel~\eqref{Eq:K}.
Depending on the level of heterogeneity within the network given by the parameter $\gamma$, 
different complex scenarios occured as the asymmetry parameter was increased, in strong
contrast to the behaviour of a classical neural field~\eqref{eq:NF} for which a single
bump exists for all $B$ with speed proportional to $B$~\cite{polngu16}.  
We found multistability, isolas of solutions, and Hopf bifurcations.
On many branches of solutions the ``twist'' of a solution increased
from zero to a maximum and then down again as the branch was traversed.
This variety of time-dependent behaviour in ``next generation'' neural field models,
as opposed to classical ones, is consistent with the observations
of others~\cite{byravi19,esnrox17,schavi19}.

Above, we have used the pulse profile $P_n(\theta)$ with $n = 2$.
However, moving bumps can be also found for other positive integers~$n$.
In particular, we have checked that the phenomena reported in this paper
also occur for $n=5$, $n=\infty$ [when $H(z;\infty)=(1-|z|^2)/(1+z+\bar{z}+|z|^2)$]
and also when synaptic dynamics are included, i.e.~replacing~\eqref{eq:I} by
\be
   \tau\frac{\partial I}{\partial t} = \int_0^{2\pi} K(x-y) H(z(y,t);n) dy - I(x,t)
\ee
where $\tau$ is the timescale of the synaptic dynamics (results not shown). Thus it seems that the scenarios
observed here are generic rather than extraordinary.

In terms of future work, it would be interesting to investigate the case of
conductance-based synaptic input~\cite{coobyr19} and opposed to the current-based
approach in~\eqref{Eq:Network}. Naturally, the existence of the complex scenarios observed
here in networks of more realistic neurons is also of interest.

{\bf Acknowledgements:} This work was initiated during a visit of CRL to the University
of Potsdam, and the hospitality of the Institute of Physics and Astronomy is acknowledged.
The work of OO was supported by the Deutsche Forschungsgemeinschaft under grant OM 99/2-1.
We thank the referees for their helpful comments.

%\bibliographystyle{plain}
%\bibliography{spiral}

\end{document}